\documentclass[10pt,twocolumn,letterpaper]{article}

\usepackage[pagenumbers]{style}

\usepackage{cuted}
\usepackage{caption}
\usepackage{graphicx}
\usepackage{amsmath,amssymb}
\usepackage{booktabs}
\usepackage{multirow}
\newcommand{\stk}[2][b]{\begin{tabular}[#1]{@{}c@{}}#2\end{tabular}}
\usepackage{balance}
\makeatletter
\newcount\km@bibcount
\AtBeginDocument{\let\km@bibitem\bibitem\def\bibitem{\global\advance\km@bibcount\@ne\ifnum\km@bibcount=72 \balance\fi\km@bibitem}}
\makeatother
\usepackage{xcolor}

\definecolor{linkblue}{rgb}{0.21,0.49,0.74}
\usepackage[pagebackref,breaklinks,colorlinks,allcolors=linkblue]{hyperref}

\title{KeyMem: Temporal Occupancy Control for Robust\\Streaming VideoLLM Memory}\author{
Wei Song\textsuperscript{1} \quad
Yuxin Cao\textsuperscript{2} \quad
Xiao Cheng\textsuperscript{3} \quad
Christy Jie Liang\textsuperscript{4} \quad
Xi Zheng\textsuperscript{3} \quad
Leo Yu Zhang\textsuperscript{1}
\\[2pt]
\textsuperscript{1}Griffith University, Australia \quad
\textsuperscript{2}National University of Singapore, Singapore
\\
\textsuperscript{3}Macquarie University, Australia \quad
\textsuperscript{4}The University of Technology Sydney, Australia
}

\def\maketitlesupplementary{\newpage\twocolumn[\centering\Large\textbf{\thetitle}\\\vspace{0.5em}Appendix\\\vspace{1.0em}]}

\begin{document}
\maketitle

\begin{abstract}
Streaming VideoLLMs retain a bounded cache of visual tokens. Yet score-based eviction lets the stream determine its own temporal occupancy, so a short salient segment can monopolize the cache and displace evidence needed later. We identify this uncontrolled temporal occupancy as a fundamental limitation and introduce KeyMem, a training-free wrapper that rank-normalizes any base score, divides the video stream into fixed-duration temporal buckets and limits how many visual tokens each bucket can retain, and optionally samples survivors with a secret key. This hard cap bounds a segment's memory share from its first chunk. On Qwen2.5-VL-7B, 20 s of burned-in captions occupies 84\% of an 8 s memory and lowers accuracy over all clips from 79.4\% to 58.3\%. Across five natural stressors, three VideoLLMs and two budgets, temporal quotas raise the worst accuracy of attention-based retention at the smaller budget from 42.2\% to 73.3--76.1\% and exceed that of InfiniPot-V by 6.1--16.7\% on two of the three models. Deterministic and keyed quotas perform comparably under natural stress. Under an adaptive adversary, a secret key preserves 4.4--9.5\% more early-clip accuracy than a leaked key, but does not consistently beat the best deterministic quota. A decide-once variant keeps old evidence that fresh draws erode. Temporal allocation is the primary robustness control, with secrecy reserved for adversarial predictability.
\end{abstract}

\section{Introduction}
\label{sec:intro}

\begin{figure*}[t]
\centering
\includegraphics[width=\textwidth]{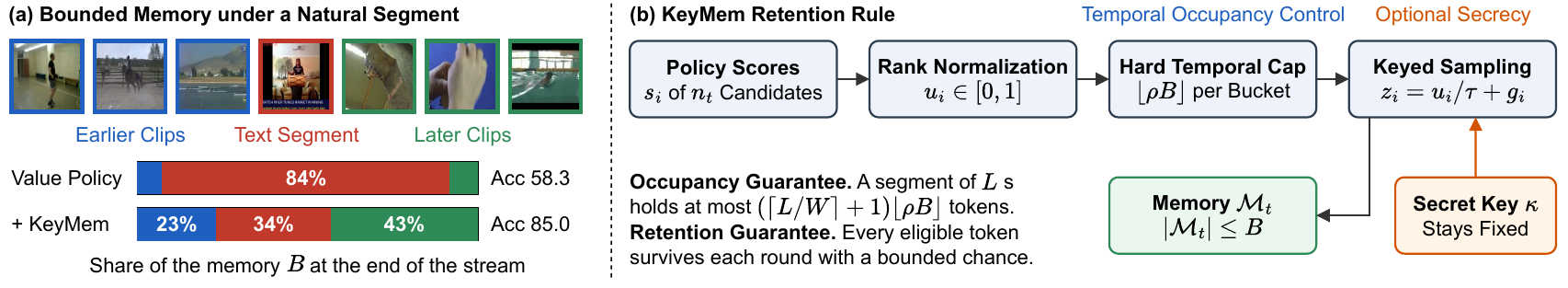}
\caption{(a) Over 30 streams of Qwen2.5-VL-7B with the value policy at $B=512$, a text segment occupies 84\% of the memory and lowers accuracy over all clips to 58.3\%, while KeyMem bounds its share and keeps 85.0\%. (b) KeyMem implements temporal occupancy control with a hard cap on the tokens of every fixed-duration bucket, which bounds the share of any segment, while equal spans controls occupancy without such a bound. Optional keyed sampling then makes the survivors unpredictable under a fixed secret key.}
\label{fig:overview}
\end{figure*}

Streaming VideoLLMs promise assistants that watch a body camera, a live broadcast or a surveillance feed and answer questions about anything that has happened so far~\cite{chen2024videollmonline,zhang2024flashvstream,qian2025dispider,wang2025streambridge}. The stream has no end, so these models keep a bounded memory of the past in the KV cache of the language model, and a retention policy evicts visual tokens after every chunk~\cite{kim2025infinipot,yang2025streammem,xu2026streamingvlm,zhang2026hermes}. Each policy scores the cached tokens by their attention, the norm of their values or their redundancy with neighbors, and it keeps the highest-scoring tokens. These scores depend on the content alone, so a short and visually busy segment can collect most of the memory. On Qwen2.5-VL-7B with a memory of 8 s of video, a 20 s segment of burned-in captions ends up holding 84\% of the memory under value-norm retention, and the accuracy over all clips falls from 79.4\% to 58.3\% (Figure~\ref{fig:overview}). The underlying problem is temporal memory concentration, which is how to stop any stretch of the stream from taking most of a fixed budget while keeping what later questions need from every part of the stream.

We call the share of the memory that a stretch of the stream holds its temporal occupancy, and temporal occupancy control regulates how the memory is spread over the periods of the stream. Score-based retention has no such control, with two consequences. Firstly, the ranking of policies does not transfer across models. Attention-based retention is among the strongest policies on Qwen2.5-VL-3B and the weakest on Qwen2.5-VL-7B. Secondly, footage that raises the scores of many tokens at once, such as captions, rapid cuts or flicker, pushes out older evidence. Adversarial exploitation is a separate and stronger motivation. Anyone who knows the model and the policy can simulate the memory offline and choose, among ordinary natural footage, the segment that erases earlier events most thoroughly, and such a segment carries no pixel perturbation that a detector could find. This raises a second question, which is whether retention that an observer cannot predict adds robustness once the temporal occupancy of every segment is bounded.

Processor caches faced both questions when Prime+Probe and Evict+Time showed that an attacker who controls which lines enter a shared cache also controls which lines leave it~\cite{osvik2006cache}. Hardware answered with partitioned caches that bound the share of any one domain~\cite{liu2016catalyst,kiriansky2018dawg} and with randomized caches that map addresses with a secret key~\cite{qureshi2018ceaser,werner2019scattercache,saileshwar2021mirage,wang2025medshield}. We carry both ideas into the token memory of a VideoLLM as two separate layers and call the result KeyMem (Figure~\ref{fig:overview}). The first layer is a hard temporal cap, which is our mechanism for temporal occupancy control. KeyMem replaces the scores of any policy by their normalized ranks and caps the tokens that any aligned time bucket can keep, so no segment of a given length holds more than a fixed share of the memory, whatever its content and whenever it plays. The second layer is optional and draws the survivors with a keyed Gumbel-top-$k$ sample, which makes them unpredictable without the key. A fresh draw in every round erodes old evidence, and a decide-once variant gives every token one draw for its whole life.

A running example separates the rules. When a 20 s caption segment follows the first clip on Qwen2.5-VL-7B, score-based retention lets it reach 89.1\% of the memory. Equal spans, which gives eight spans of the elapsed stream equal shares, is another form of temporal occupancy control, but it sets no upper bound, and the segment holds 68.5\% of the young stream's memory. Cap only, which applies the cap without the keyed draw, and KeyMem hold it at 37.5\% from the first chunk but keep no more earlier evidence, since equal spans answers 91.1\% of the questions about the earlier clips, cap only 86.7\% and KeyMem 82.2\% (Section~\ref{sec:burst}). The cap is thus a guarantee on memory share, the key is a guarantee on predictability, and neither guarantees the accuracy of an answer.

We evaluate these rules against deterministic temporal quotas and two published retention rules, InfiniPot-V~\cite{kim2025infinipot} and Nexus Sampling~\cite{duong2026nexus}, on a coverage benchmark that asks about every clip of 80 s streams built from UCF101~\cite{soomro2012ucf101}. The streams carry five natural stress segments and two common corruptions, and we test three VideoLLMs at two budgets with three base policies. We extend the streams to 400 s, move a burst through the stream, and test an adaptive adversary that knows the quota and searches up to 64 natural choices over several keys, with full memory as a control for direct interference. The results show that temporal allocation removes the concentration failures and that the key helps only against an adaptive adversary.

Our contributions are as follows.
\begin{itemize}
\item We identify uncontrolled temporal occupancy as a fundamental limitation of score-based retention and show that temporal allocation removes its failures across models, retention policies and natural stressors.
\item We introduce a hard temporal occupancy cap and prove that it bounds the memory share of any fixed-duration segment from its first chunk, whatever the content, which equal spans does not.
\item We add optional secret-keyed sampling, which makes retention less predictable and helps against an adaptive adversary, and decide-once sampling, which limits the erosion of old evidence on long streams.
\end{itemize}

\section{Related Work}
\label{sec:related}

\noindent\textbf{Streaming VideoLLMs and Memory.} Online VideoLLMs keep a bounded summary of the past~\cite{chen2024videollmonline,zhang2024flashvstream,zhang2025flashvstream,qian2025dispider,wang2025streambridge}, and many recent systems keep it in the KV cache of the language model. InfiniPot-V~\cite{kim2025infinipot} compresses the cache with temporal redundancy and value norms, StreamMem~\cite{yang2025streammem} scores tokens with generic query tokens, and StreamingVLM~\cite{xu2026streamingvlm} keeps attention sinks and a short window. HERMES~\cite{zhang2026hermes}, StreamingTOM~\cite{chen2026streamingtom}, MuKV~\cite{xiao2026mukv}, LiveVLM~\cite{ning2026livevlm} and Going Down Memory Lane~\cite{agarwal2026memorylane} refine the scoring, the granularity or the layout of this memory, ReKV~\cite{di2025rekv} and V-Rex~\cite{kim2026vrex} retrieve from an offloaded cache, and TimeChat-Online~\cite{yao2025timechatonline} drops redundant tokens before they reach it. ProtoKV~\cite{minh2026protokv} notes that a brief decisive cue can be evicted long before a question asks about it. DeltaS~\cite{kwon2026deltas} gives equal spans of elapsed time an equal share of the budget, but unused shares go to the highest-scoring tokens anywhere, so no span has a hard upper limit. These methods choose the survivors deterministically and report average accuracy on OVO-Bench~\cite{niu2025ovobench}, StreamingBench~\cite{lin2024streamingbench} and RVS~\cite{zhang2024flashvstream}.

\noindent\textbf{KV Cache Eviction and Its Robustness.} Text LLMs evict KV entries with sink tokens and a recent window~\cite{xiao2024streamingllm}, accumulated attention~\cite{zhang2023h2o,liu2023scissorhands}, an observation window~\cite{li2024snapkv} and budgets per head or per layer~\cite{ge2024fastgen,cai2025pyramidkv}, and visual token reduction follows the same pattern~\cite{chen2024fastv,tao2025dycoke,huang2025prunevid,shen2025fastvid,yang2025visionzip,shen2025longvu,bolya2023tome}. DefensiveKV~\cite{feng2026defensivekv} and ReST-KV~\cite{an2026restkv} make importance estimates robust to drift, and Nexus Sampling~\cite{duong2026nexus} replaces top-$k$ selection with weighted reservoir sampling, so that every token with a positive weight can survive each step. Stochastic eviction also serves cache diversity~\cite{chang2026vase}, matches scored eviction on reasoning~\cite{wang2026randomattention}, certifies the eviction error~\cite{xie2026certificates} and estimates expectations by sampling~\cite{geh2026probabilistic}, and a structural reserve of the cache can matter more than the scorer~\cite{garcia2026protection}. None of these works keys its randomness or bounds the share of a stretch of time. On the security side, RobustKV~\cite{jiang2025robustkv} evicts jailbreak content, Hossain \etal~\cite{hossain2025corrupted} corrupt cached keys, Dobre \etal~\cite{dobre2026visualmemory} show that an adversarial image can act through the cache after it has left the context, and Song \etal~\cite{song2026kvprovenance} show that shared caches lose the provenance of their entries.

\begin{table}[t]
\centering
\footnotesize
\caption{Comparison with related retention rules.}
\label{tab:novelty}
\setlength{\tabcolsep}{1.6pt}
\begin{tabular}{@{}llcccc@{}}
\toprule
Method & Domain & \stk{Temporal\\Quota} & \stk{Survival\\Bound} & \stk{Secret\\Key} & \stk{Threat\\Evaluation} \\
\midrule
H2O, SnapKV & Text & None & None & No & No \\
InfiniPot-V & Video & None & None & No & No \\
DeltaS & Video & Soft & None & No & No \\
Nexus & Text & None & Per step & No & No \\
VaSE~\cite{chang2026vase} & Text & None & None & No & No \\
Random attention~\cite{wang2026randomattention} & Text & None & None & No & No \\
Error certificates~\cite{xie2026certificates} & Text & None & Error & No & No \\
KeyMem (ours) & Video & Hard & Per round & Yes & Yes \\
Decide-once (ours) & Video & Hard & Fixed ranks & Yes & Yes \\
\bottomrule
\end{tabular}
\end{table}

\noindent\textbf{Comparison.} Table~\ref{tab:novelty} contrasts our rules with the closest retention rules. DeltaS sets a soft quota over elapsed time, and our cap sets a hard limit for a fixed duration. Nexus states a per-step survival property for a public sampler, and none of the compared works keys its randomness or evaluates content chosen by simulating the rule.

\noindent\textbf{Robustness of Video Models.} Benchmarks of natural corruptions extend from ImageNet-C~\cite{hendrycks2019benchmarking} to action recognition~\cite{yi2021benchmarking,schiappa2023largescale}, video-language models~\cite{schiappa2022robustness}, vision-language adaptation~\cite{chen2023benchmarking} and multimodal LLMs~\cite{zhang2024mmcbench,agarwal2025mvtamperbench,wang2025trustvideollms}\,{\cite{fan2026server}}, and they measure an offline model under a perturbed input. Adversarial work shows that the temporal pipeline of video models is itself an attack surface, from style perturbations against video classifiers~\cite{cao2023stylefool,cao2024logostylefool} and defenses of video recognition~\cite{song2024secvid,song2025vidtoken}\,{\cite{chen2023fedward}} to frame sampling that hides harmful content from VideoLLMs~\cite{cao2026failures,cao2025poisonvid}\,{\cite{long2024fedcl}}, stress tests that provoke degenerate repetition~\cite{cao2026videostf} and temporal sampling that gates claim deference~\cite{cao2026caved}. We borrow two corruptions from these benchmarks and measure what a bounded memory forgets about the rest of a stream.

\section{Problem Setting}
\label{sec:setting}

\noindent\textbf{Streaming Memory.} A streaming VideoLLM receives frames in chunks of a few seconds and never sees the end of the stream. After each chunk, the language model appends the keys and values of the new visual tokens to its KV cache, which is the only memory of the past. We write $\mathcal{M}_t$ for the set of visual tokens held after chunk $t$ and $B$ for the token budget. System and text tokens are never removed and do not count toward $B$. When appending chunk $t$ brings the number of candidates $n_t=|\mathcal{M}_{t-1}|+c_t$ above $B$, where $c_t$ is the number of new tokens, a retention policy chooses which $B$ tokens survive. Every removed token is gone for good, because the stream is not stored.

\noindent\textbf{Score-Based Retention.} Most retention policies assign each candidate token $i$ a score $s_i$ and keep the $B$ highest. These scores come from attention~\cite{zhang2023h2o,li2024snapkv}, value norms~\cite{kim2025infinipot} or similarity with neighbors~\cite{bolya2023tome,yao2025timechatonline,tao2025dycoke}, while a sliding window keeps the newest tokens and ignores content. All of these rules are deterministic functions of the content, so busy footage such as rapid cuts, burned-in captions, flicker, camera shake and fast motion, which raises the scores of many tokens at once, also raises their share of the memory and can push out older evidence that a later question needs.

\noindent\textbf{Threat Model.} Busy footage can also be chosen on purpose. We consider an adversary who places natural footage into the stream, for example by holding a screen in view of a body camera or by inserting a clip into a broadcast, and who wants the model to forget what happened before. Following Kerckhoffs's principle, the adversary knows the model, the base policy, the budget, the quota, the bucket grid and every hyperparameter, and it chooses both the content of its segment and the moment at which the segment starts. It also knows the earlier footage and the questions, and its candidates never show a class of the stream or of the answer options. It can simulate the memory offline, but it cannot edit pixels after capture, query the deployed system or observe its answers. The simulation is exact against a deterministic policy, a randomized policy with a public seed and a keyed policy with a leaked key, and against a secret key, the adversary can only simulate with keys of its own.

\noindent\textbf{Goal.} We want a training-free rule that wraps any score-based policy, runs at negligible cost, and bounds for every content how much of the memory any short segment can occupy. A secondary and optional goal is retention that an observer without the key cannot predict, together with a survival bound for every token that passes the cap.

\section{Temporal Occupancy and Keyed Retention}
\label{sec:method}

Our retention rule transfers the keyed eviction of randomized caches~\cite{qureshi2018ceaser,qureshi2019ceasers,werner2019scattercache,saileshwar2021mirage,wang2025privacy} and the quotas of partitioned caches~\cite{liu2016catalyst,kiriansky2018dawg} to the token memory of a VideoLLM.

\noindent\textbf{Rank Normalization.} The raw scores of different policies live on different scales, and a few extreme scores can dominate any rule that uses them directly. We therefore replace every score by its normalized rank among the $n_t$ candidates,
\begin{equation}
u_i=\frac{\operatorname{rank}(s_i)}{n_t-1}\in[0,1],
\label{eq:rank}
\end{equation}
where ties receive their mean rank. A burst of inflated scores thus gains no more weight than the top rank.

\noindent\textbf{Hard Temporal Cap.} KeyMem controls temporal occupancy with a hard cap. We split the timeline into aligned buckets of $W$ seconds and allow each bucket to hold at most $\lfloor\rho B\rfloor$ tokens, where $\rho\in(0,1]$ is the cap ratio. Within a bucket that exceeds the cap, only the $\lfloor\rho B\rfloor$ tokens with the highest $u_i$ remain eligible. The cap is applied after every chunk, including chunks after which the memory is still under budget, so it holds at all times.

\noindent\textbf{Keyed Sampling.} Among the eligible set $\mathcal{E}_t$, we keep $m_t=\min(B,|\mathcal{E}_t|)$ tokens with the Gumbel-top-$k$ trick~\cite{kool2019stochastic}. Each eligible token receives a perturbed score
\begin{equation}
z_i=\frac{u_i}{\tau}+g_i,\qquad g_i=-\log\bigl(-\log\xi_i\bigr),
\label{eq:gumbel}
\end{equation}
where $\tau>0$ is a temperature and $\xi_i$ is a uniform variable drawn from a pseudorandom generator seeded by a hash of a secret key $\kappa$, the epoch $\lfloor t/R\rfloor$ of $R$ chunks and the chunk index $t$. The $m_t$ tokens with the largest $z_i$ survive. Kool \etal~\cite{kool2019stochastic} show that this selection has the same distribution as drawing $m_t$ tokens without replacement with probabilities proportional to the weights $w_i=e^{u_i/\tau}$, which is the Plackett-Luce model~\cite{plackett1975analysis,luce1959individual}. Because $u_i\in[0,1]$, every weight lies in $[1,e^{1/\tau}]$, so the temperature directly sets the largest advantage that any token can have over any other. A small $\tau$ follows the base policy closely, and a large $\tau$ approaches uniform random retention. The seed changes in every round under one fixed key, and an observer who does not hold the key cannot predict which tokens survive.

\noindent\textbf{Occupancy Guarantee.} The cap gives a deterministic bound on how much memory any segment can take, which Proposition~1 states. \emph{Proposition 1. For any segment of the stream that lasts $L$ seconds, the number of its tokens in $\mathcal{M}_t$ never exceeds $(\lceil L/W\rceil+1)\lfloor\rho B\rfloor$, at every time $t$ and for every content of the stream.} The segment meets at most $\lceil L/W\rceil+1$ aligned buckets, and each bucket holds at most $\lfloor\rho B\rfloor$ tokens after every chunk. The bound holds whatever the base scores and the random draws are, so at least $B-(\lceil L/W\rceil+1)\lfloor\rho B\rfloor$ of the $B$ slots stay closed to the segment, although other content need not fill them.

\noindent\textbf{Retention Guarantee.} The bounded weight ratio also gives every eligible token a chance to survive each round, however low its score is, as Proposition~2 states. \emph{Proposition 2. Let $a=e^{1/\tau}$, let $n=|\mathcal{E}_t|$ eligible tokens compete for $m=m_t$ slots, and let $r=n-m$ be the number of evictions. For every eligible token $i$ and every assignment of scores, the probability that $i$ is evicted in round $t$ satisfies}
\begin{equation}
\Pr[i\notin\mathcal{M}_t]\le\prod_{j=r}^{n-1}\frac{ja}{1+ja}\le\Bigl(\frac{1+ar}{1+an}\Bigr)^{1/a}.
\label{eq:retention}
\end{equation}

\noindent The bound follows from the sequential view of the Plackett-Luce draw, in which the draw after $k$ earlier draws selects token $i$ with probability at least $1/(1+(n-k)a)$ while $i$ is still unselected (Appendix~\ref{sec:proofs}). As $\tau\to\infty$, the product telescopes to $r/n$, the eviction rate of uniform random retention. With $B=512$, $\lfloor\rho B\rfloor=96$ and buckets of 10 s, a chunk of 4 s meets at most two buckets and adds at most 192 eligible tokens, so at most 704 tokens compete for 512 slots, and every eligible token survives each round with a probability of at least 16.1\%, whatever its score.

\noindent\textbf{Decide-Once Sampling.} The fresh draw of every round has a cost over long streams, because a token that survives one round faces eviction again in the next. We therefore also study a decide-once variant, which keeps the ranks, the cap and the perturbed score of Equation~\ref{eq:gumbel} but takes $\xi_i$ from a hash of the key and the identifier of token $i$, so that each token holds one perturbation for its whole life. Proposition~3 states the effect when the ranks do not change. \emph{Proposition 3. Let the ranks stay fixed, let $\mathcal{A}_t$ be the set of tokens that the cap has left eligible in at least one round up to $t$, and let $N=|\mathcal{A}_t|$. Under decide-once sampling, every token that the cap keeps eligible through round $t$ satisfies Equation~\ref{eq:retention} for its eviction by round $t$, with $n=N$ and $m=\min(B,N)$.} One draw over $\mathcal{A}_t$ thus replaces a separate draw in every round (Appendix~\ref{sec:proofs}). On a 200 s stream, where each of 20 buckets offers 96 tokens, the bound gives a survival probability of at least 4.1\%, while the per-round bounds of KeyMem multiply to below $10^{-5}$.

\noindent\textbf{Scope of the Guarantees.} The propositions bound the memory, and none of them bounds the accuracy of an answer. Proposition 1 guarantees the earlier clips neither tokens nor accuracy, and it becomes vacuous once $(\lceil L/W\rceil+1)\lfloor\rho B\rfloor\ge B$. Proposition 2 holds only for the tokens that the cap leaves eligible, so a token that ranks below the cap of an overfull bucket is evicted for certain. It also holds per round, so over $k$ rounds, the survival of a token is bounded below only by a product of $k$ per-round bounds, which decays geometrically (Section~\ref{sec:long}). Proposition 3 assumes fixed ranks, which real policies do not provide, so it only explains the mechanism. Proposition 2 also does not carry over to later rounds of decide-once, because the survivors of earlier rounds hold larger perturbations. All probabilities treat the keyed generator as a random function of the key and assume that the content does not depend on the key. Appendix~\ref{sec:proofs} gives the proofs.

\noindent\textbf{Comparison of Temporal Guarantees.} A hard cap, equal spans and a soft quota are three forms of temporal occupancy control, and they promise different things. The hard cap of Proposition 1 bounds a segment by its length alone, from the first chunk and at every elapsed time, so the share of a burst is known before the burst plays, whatever its content, its timing and the length of the stream. Equal spans keeps a share of $B/8$ for every old span with enough tokens, and this floor persists at any stream length. A span grows with the elapsed time, however, so a burst in a young stream can fill most of the memory, and an adversary who chooses when its segment starts also chooses its share. The soft quota of DeltaS~\cite{kwon2026deltas} also allocates by elapsed time, and its unused shares go to the highest-scoring tokens anywhere, so no span has an upper limit. The cap pays for its bound with retention. It gives old evidence no floor and also caps the buckets of that evidence, so it can keep fewer earlier tokens than equal spans (Section~\ref{sec:burst}). The keyed draw only changes which tokens survive inside the quota, so it can matter only against an adversary who simulates the memory, which Section~\ref{sec:adversary} tests.

\section{Experiments}
\label{sec:experiments}

Following the three contributions, Sections~\ref{sec:main} to~\ref{sec:sbench} measure the benefit of temporal allocation, Section~\ref{sec:burst} tests the guarantee of the hard cap, and Sections~\ref{sec:adversary} and~\ref{sec:long} measure when keyed sampling helps.

\subsection{Experimental Setup}
\label{sec:setup}

\noindent\textbf{Streams and Questions.} Every stream concatenates the first 10 s of six clips from six action classes of the UCF101 test split~\cite{soomro2012ucf101}, sampled at 2 frames per second, and a 20 s stress segment follows the third clip, so every stream lasts 80 s. After the stream ends, the model answers one question per clip, which asks which of five actions was performed in the video. The four wrong options appear nowhere in the stream, so a correct answer requires memory of the clip, and circular evaluation averages the score over the five cyclic orders of the options. We evaluate 30 test streams.

\begin{figure}[t]
\centering
\includegraphics[width=\columnwidth]{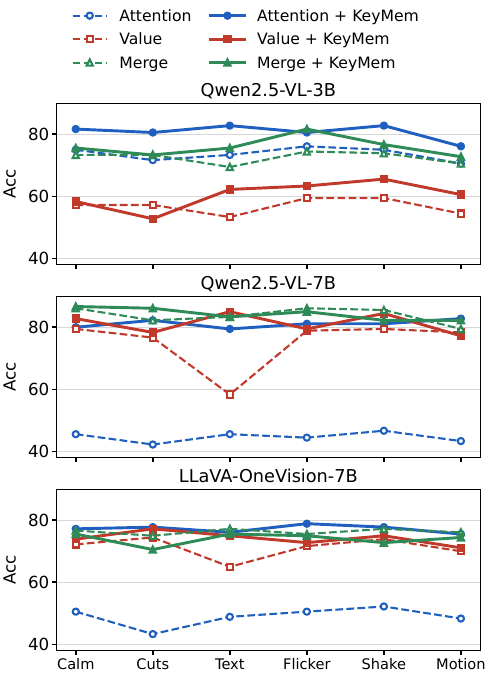}
\caption{Accuracy per stress type.}
\label{fig:stress}
\end{figure}

\noindent\textbf{Stress Segments.} The calm control plays the calmest tenth of the test clips by frame difference, motion plays the busiest tenth, and cuts plays twenty 1 s shots. Text overlays the calm footage with captions, a ticker and a clock, flicker alternates the brightness of its frames, and shake rotates, shifts and zooms them. Gaussian noise and blur from robustness benchmarks~\cite{hendrycks2019benchmarking} corrupt the calm footage, and we report them apart from these five stressors. No segment is optimized against a model or a policy, and Figure~\ref{fig:stress} gives the accuracy under each at the 8 s budget.

\begin{table*}[t]
\centering
\small
\caption{Comparison with temporal quotas.}
\label{tab:quota}
\begin{tabular}{@{}llcccccccc@{}}
\toprule
\multirow{2}{*}{Model} & \multirow{2}{*}{Policy} & \multicolumn{2}{c}{Base policy} & \multicolumn{2}{c}{+ Equal spans} & \multicolumn{2}{c}{+ Cap only} & \multicolumn{2}{c}{+ KeyMem} \\
\cmidrule(lr){3-4}\cmidrule(lr){5-6}\cmidrule(lr){7-8}\cmidrule(lr){9-10}
 & & Calm & Worst Acc & Calm & Worst Acc & Calm & Worst Acc & Calm & Worst Acc \\
\midrule
\multirow{3}{*}{Qwen2.5-VL-3B} & Attention & 75.0 & 70.6 & 82.2 & 73.3 & 82.2 & \textbf{77.8} & 81.7 & 76.1 \\
 & Value & 57.2 & 53.3 & 58.3 & 53.3 & 59.4 & \textbf{55.6} & 58.3 & 52.8 \\
 & Merge & 73.3 & 69.4 & 73.3 & \textbf{73.3} & 75.0 & 68.3 & 75.6 & 72.8 \\
\midrule
\multirow{3}{*}{Qwen2.5-VL-7B} & Attention & 45.6 & 42.2 & 83.3 & \textbf{80.0} & 78.3 & 78.3 & 80.0 & 79.4 \\
 & Value & 79.4 & 58.3 & 82.8 & \textbf{79.4} & 81.7 & 75.0 & 82.8 & 77.2 \\
 & Merge & 86.1 & 79.4 & 86.7 & 81.7 & 85.6 & \textbf{83.9} & 86.7 & 82.2 \\
\midrule
\multirow{3}{*}{LLaVA-OneVision-7B} & Attention & 50.6 & 43.3 & 76.7 & 74.4 & 79.4 & \textbf{76.1} & 77.2 & 75.6 \\
 & Value & 72.2 & 65.0 & 74.4 & 70.0 & 72.8 & \textbf{71.1} & 73.9 & \textbf{71.1} \\
 & Merge & 76.7 & \textbf{75.0} & 77.2 & 74.4 & 75.6 & 74.4 & 75.6 & 70.6 \\
\bottomrule
\end{tabular}
\end{table*}

\noindent\textbf{Models and Policies.} We evaluate Qwen2.5-VL-3B, Qwen2.5-VL-7B~\cite{bai2025qwen25vl} and LLaVA-OneVision-7B~\cite{li2025llavaonevision} with chunks of 4 s. The budget $B$ holds 8 s or 16 s of video, which is $B=512$ or $B=1024$ on Qwen2.5-VL and $B=3136$ or $B=6272$ on LLaVA-OneVision, and full memory without a budget gives a reference. The attention policy scores a token by the attention that the newest chunk pays to it in the last layer~\cite{zhang2023h2o,li2024snapkv}, the value policy by its value norm~\cite{kim2025infinipot}, and the merge policy by its difference from the previous token~\cite{bolya2023tome,tao2025dycoke}. KeyMem wraps each of them with $\tau=0.5$, $R=8$, $W=10$ s and $\rho=0.1875$, which 12 validation streams from the training split select (Appendix~\ref{app:validation}). Equal spans gives 8 equal spans of the elapsed stream an equal share of the budget, in the style of DeltaS~\cite{kwon2026deltas}, cap only sets $\tau=0.01$, and decide-once keeps one perturbation per token. We also run InfiniPot-V, Nexus Sampling~\cite{duong2026nexus}, a sliding window and random retention with uniform weights.

\noindent\textbf{StreamingBench.} We also stream the 150 videos of the real-time visual understanding task of StreamingBench~\cite{lin2024streamingbench} on Qwen2.5-VL-7B at 1 frame per second, with $B=2048$ for about 28 s of video, and answer each of its 721 questions from the memory at its own time.

\noindent\textbf{Adversaries.} On all 30 test streams of Qwen2.5-VL-7B at $B=512$, the adversary offers 16 natural candidate segments of 20 s at two start times and scores the 32 choices by simulating the memory. The token objective counts the surviving tokens of the three earlier clips, and the answer objective sums the margin of the correct option on their questions. Keyed defenses run with three defender keys, and the adversary simulates with three keys of its own.

\noindent\textbf{Metrics.} Accuracy (Acc) is the fraction of clip questions answered correctly, and early accuracy (Early Acc) is the fraction over the three clips before the stress segment. Residency (Res) is the fraction of the visual tokens of a clip that are still in memory at the questions. Worst-case accuracy (Worst Acc) is the lowest accuracy over models or stress types, and Excess is the drop in Early Acc of a defense under an attack minus that of full memory on the same segments. Accuracy and residency are given in \%, brackets give 95\% paired bootstrap intervals over streams, and Appendix~\ref{app:impl} gives the full setup.

\begin{table}[t]
\centering
\caption{Comparison with published retention rules.}
\label{tab:sota}
\setlength{\tabcolsep}{2.5pt}
\resizebox{\columnwidth}{!}{%
\begin{tabular}{@{}lccccccccc@{}}
\toprule
\multirow{3}{*}{Rule} & \multicolumn{3}{c}{Qwen2.5-VL-3B} & \multicolumn{3}{c}{Qwen2.5-VL-7B} & \multicolumn{3}{c}{LLaVA-OneVision-7B} \\
\cmidrule(lr){2-4}\cmidrule(lr){5-7}\cmidrule(lr){8-10}
& Calm & \stk{Worst\\Acc} & Corruption & Calm & \stk{Worst\\Acc} & Corruption & Calm & \stk{Worst\\Acc} & Corruption \\
\midrule
Full memory & 84.4 & 80.6 & 84.4 & 90.6 & 83.3 & 88.6 & 86.7 & 80.0 & 86.9 \\
InfiniPot-V & 70.6 & 60.0 & 67.2 & 70.6 & 67.2 & 70.8 & 70.6 & 63.9 & 73.6 \\
Nexus Sampling & 37.2 & 32.8 & 36.9 & 43.3 & 38.3 & 42.8 & 46.7 & 43.3 & 46.1 \\
Base policies & 68.5 & 64.4 & 68.6 & 70.4 & 60.0 & 71.5 & 66.5 & 61.1 & 68.0 \\
Equal spans & 71.3 & 66.7 & 72.9 & 84.3 & 80.4 & 84.2 & 76.1 & 73.0 & 79.4 \\
Cap only & 72.2 & 67.2 & 72.2 & 81.9 & 79.1 & 82.7 & 75.9 & 73.9 & 75.8 \\
KeyMem & 71.9 & 67.2 & 74.9 & 83.1 & 79.6 & 82.9 & 75.6 & 72.4 & 76.1 \\
Decide-once & 73.9 & 70.0 & 74.0 & 83.5 & 80.9 & 82.4 & 75.9 & 73.9 & 76.8 \\
\bottomrule
\end{tabular}}
\end{table}

\subsection{Natural Stress}
\label{sec:main}

Table~\ref{tab:quota} compares the rules on all 30 test streams at the 8 s budget, and Table~\ref{tab:main} in Appendix~\ref{app:full} adds the 16 s budget and the reference policies. Firstly, the ranking of base policies does not transfer across models. Attention is the best base policy on Qwen2.5-VL-3B with 75.0\% in calm streams, but it falls to 45.6\% on Qwen2.5-VL-7B and to 50.6\% on LLaVA-OneVision, close to the sliding window. Secondly, bounding the temporal occupancy removes the collapses, whichever quota does it. The Worst Acc of attention rises from 42.2\% on Qwen2.5-VL-7B to 80.0\% with equal spans, 78.3\% with cap only and 79.4\% with KeyMem. Averaged over the nine combinations of model and base policy, the Worst Acc is 73.3\% with equal spans, 73.4\% with cap only and 73.1\% with KeyMem, against 61.8\% for the base policies, and the 18 paired intervals of KeyMem against the two deterministic quotas all include zero (Appendix~\ref{app:boot}). The protection thus comes from the bound on occupancy, and the keyed draw adds nothing measurable on natural stress. Thirdly, a quota does not repair a base policy that is weak on a model. Value on Qwen2.5-VL-3B stays between 52.8\% and 55.6\% with every quota, and KeyMem lowers the Worst Acc of merge on LLaVA-OneVision from 75.0\% to 70.6\%. Without the cap, the keyed draw protects no old evidence (Appendix~\ref{app:ablation}), and a retention decision takes at most 13.3 ms on one CPU core (Appendix~\ref{app:overhead}).

Figure~\ref{fig:stress} shows that most stressors move the accuracy of a policy only slightly. The collapse of attention on Qwen2.5-VL-7B and LLaVA-OneVision-7B appears under every stress type and in calm streams alike, and KeyMem lifts attention to the level of the other policies on both models. The text segment under value is the exception of Figure~\ref{fig:overview}. With KeyMem, the segment holds 34\% of the memory, and the accuracy stays at 85.0\%, and since the segment reuses the calm footage, the whole drop of the base policy comes from burned-in text.

\subsection{Published Retention Rules}
\label{sec:sota}

Table~\ref{tab:sota} compares the temporal rules with InfiniPot-V and Nexus Sampling at the 8 s budget, with the quotas averaged over the three base policies and the Corruption column averaging the accuracy under noise and blur. InfiniPot-V is the stronger published rule, since its temporal-redundancy share keeps tokens from every clip. It trails all nine combinations of quota and base policy in Worst Acc on Qwen2.5-VL-7B, by 7.8\% to 16.7\% with every interval above zero, and on LLaVA-OneVision-7B, by 6.1\% to 12.2\% with six intervals above zero and the other three ending at zero. On Qwen2.5-VL-3B, it trails every quota with attention or merge, by 8.3\% to 17.8\%, and leads those built on value, the weakest base policy of that model, by 4.4\% to 7.2\% (Table~\ref{tab:sotaci} in Appendix~\ref{app:sota}).
Nexus Sampling weights blocks by their attention mass, which in a video falls mostly on the newest chunk, so it reaches a Worst Acc of only 32.8\% to 43.3\%. At the 16 s budget, InfiniPot-V reaches 70.0\% to 72.8\% and Nexus Sampling 50.6\% to 63.3\%, all below KeyMem with attention at 77.8\% to 80.6\%. Under the corruptions, every quota leads InfiniPot-V by 10.6\% to 14.7\% on Qwen2.5-VL-7B, with every interval above zero, and by -0.3\% to 7.5\% on LLaVA-OneVision-7B, and it leads Nexus Sampling by at least 27.2\% on both models. Neither published rule bounds the share of a segment, and Nexus Sampling draws from a public seed, so an adversary can simulate either rule exactly.

\subsection{Streaming Question Answering}
\label{sec:sbench}

\begin{table}[t]
\centering
\caption{Peak share and early accuracy when a text segment moves through the stream, averaged over the three base policies, with $^{\dagger}$ for a shifted grid.}
\label{tab:burst}
\footnotesize
\setlength{\tabcolsep}{2.4pt}
\begin{tabular}{@{}lcccccccc@{}}
\toprule
\multirow{3}{*}{\begin{tabular}{@{}l@{}}Segment\\(s)\end{tabular}} & \multirow{3}{*}{Bound} & \multicolumn{3}{c}{Peak share} & \multicolumn{4}{c}{Early Acc} \\
\cmidrule(lr){3-5}\cmidrule(lr){6-9}
& & Base & \stk{Equal\\spans} & \stk{Cap\\only} & \stk{Equal\\spans} & \stk{Cap\\only} & KeyMem & \stk{Decide-\\once} \\
\midrule
10--30 & 37.5 & 89.1 & 68.5 & 37.5 & \textbf{91.1} & 86.7 & 82.2 & 90.0 \\
20--40 & 37.5 & 84.6 & 50.8 & 37.5 & \textbf{85.0} & 74.4 & 75.0 & 83.9 \\
30--50 & 37.5 & 79.3 & 39.7 & 37.5 & \textbf{84.8} & 80.0 & 77.8 & 81.5 \\
50--70 & 37.5 & 74.3 & 27.1 & 35.9 & \textbf{82.7} & 77.1 & 77.6 & 78.9 \\
10--30$^{\dagger}$ & 56.2 & 89.1 & 68.5 & 54.0 & 91.1 & 87.8 & 90.0 & \textbf{92.2} \\
20--40$^{\dagger}$ & 56.2 & 84.6 & 50.8 & 54.2 & \textbf{85.0} & 72.8 & 75.6 & 83.9 \\
10--50 & 75.0 & 95.3 & 80.0 & 75.0 & \textbf{90.0} & 74.4 & 74.4 & 82.2 \\
\bottomrule
\end{tabular}
\end{table}

\begin{table*}[t]
\centering
\footnotesize
\caption{Early accuracy under the adaptive adversary on Qwen2.5-VL-7B at $B=512$, with $^{*}$ for an excess whose interval lies above zero.}
\label{tab:adaptive}
\setlength{\tabcolsep}{3.5pt}
\begin{tabular}{@{}lcccccccccccc@{}}
\toprule
\multirow{2}{*}{Defense} & \multicolumn{4}{c}{Attention} & \multicolumn{4}{c}{Value} & \multicolumn{4}{c}{Merge} \\
\cmidrule(lr){2-5}\cmidrule(lr){6-9}\cmidrule(lr){10-13}
 & Calm & Token & Answer & Excess & Calm & Token & Answer & Excess & Calm & Token & Answer & Excess \\
\midrule
Full memory & 86.7 & -- & 64.4 & -- & 86.7 & -- & 64.4 & -- & 86.7 & -- & 64.4 & -- \\
\midrule
Base policy & 27.8 & 25.6 & 22.2 & 0.0 & 73.3 & 56.7 & 45.6 & 17.8$^{*}$ & \textbf{87.8} & 80.0 & 71.1 & 1.1 \\
Equal spans & \textbf{85.6} & \textbf{83.3} & \textbf{75.6} & -3.3 & \textbf{80.0} & 76.7 & 58.9 & 8.9 & 84.4 & 78.9 & 71.1 & -3.3 \\
Cap only & 70.0 & 65.6 & 46.7 & 12.2$^{*}$ & 77.8 & 75.6 & 61.1 & 3.3 & \textbf{87.8} & \textbf{84.4} & 72.2 & -1.1 \\
KeyMem, leaked key & 71.5 & 72.6 & 47.8 & 11.9$^{*}$ & 75.2 & 74.8 & 52.2 & 10.4$^{*}$ & 83.0 & 79.6 & 67.4 & 0.7 \\
KeyMem, secret key & 71.5 & 72.6 & 56.2 & 1.6 & 75.2 & 74.3 & 61.7 & 0.1 & 83.0 & 82.0 & 71.9 & -3.7 \\
Decide-once, leaked key & 77.0 & 79.3 & 58.9 & 3.7 & 79.6 & 81.9 & 63.0 & 3.0 & 82.6 & 79.6 & 70.4 & -2.6 \\
Decide-once, secret key & 77.0 & 77.3 & 61.1 & 2.2 & 79.6 & \textbf{83.2} & \textbf{67.3} & -2.5 & 82.6 & 82.1 & \textbf{73.8} & -4.6 \\
\bottomrule
\end{tabular}
\end{table*}
Table~\ref{tab:sbench} in Appendix~\ref{app:sbench} reports the accuracy on StreamingBench over all 721 questions and over the 414 questions asked after 180 s. Full memory reaches 69.3\% and the sliding window 64.5\%, so a memory of 28 s loses 4.8\%. No interval of a quota against its base policy lies below zero, KeyMem raises value by 3.2\% [0.6, 5.9], and decide-once shows no clear gain over its base policies. Over all questions, equal spans raises attention from 66.0\% to 67.8\%, and KeyMem raises merge from 65.5\% to 67.3\%. The gap to full memory is widest after 180 s, where the quotas differ most. Temporal allocation thus also helps on real questions, where no segment is designed to stress the memory.

\subsection{Burst Placement}
\label{sec:burst}

To test the guarantee that only the hard cap gives, we move the 20 s text segment through the 80 s streams of Qwen2.5-VL-7B at $B=512$, shift the bucket grid by 5 s and lengthen the segment to 40 s. The peak share is the largest fraction of the budget that the segment holds after any chunk, and Early Acc covers every clip before the segment. \mbox{Table}~\ref{tab:burst} shows the trade-off that the cap makes. Firstly, the cap limits concentration. The capped rules never exceed the bound of Proposition 1 for the buckets that the segment meets, while equal spans lets the segment hold 68.5\% of the memory after the first clip and 27.1\% after the fifth, and the base policies reach 74.3\% to 95.3\%. Secondly, the cap keeps no more earlier evidence than equal spans. Each earlier clip fills one bucket of 640 tokens, of which the cap keeps 96, so cap only and KeyMem trail equal spans on Early Acc in all seven placements, by 3.3\% to 15.6\% and 1.1\% to 15.6\%, with 11 of their 14 paired intervals below zero (Appendix~\ref{app:burst}). Limiting concentration and keeping earlier evidence are thus separate properties. This separation is a finding of its own, since a guarantee on the share of a burst gives no guarantee on the evidence that later questions need. The cap is worth its cost when the share of a burst must be known from its first chunk, since the start time alone moves the share under equal spans from 27.1\% to 68.5\%.

\subsection{Adaptive Adversary}
\label{sec:adversary}

\begin{figure}[t]
\centering
\includegraphics[width=0.96\columnwidth]{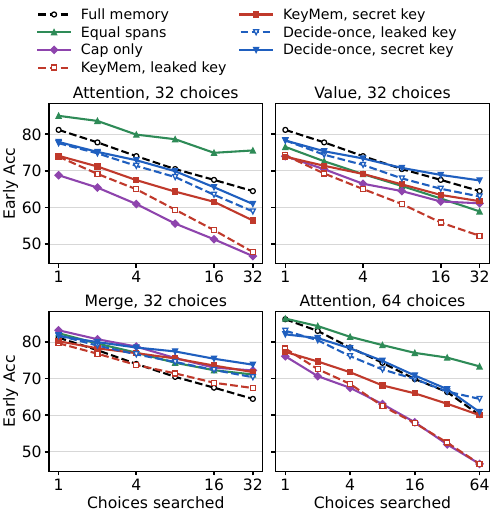}
\caption{Early accuracy against the answer attack as the search grows, averaged over random subsets of the choices.}
\label{fig:adaptive}
\end{figure}

Table~\ref{tab:adaptive} reports the Early Acc of every defense against the search of all 32 choices, averaged over the key pairs, while Figure~\ref{fig:adaptive} follows the damage as the search grows, and Appendix~\ref{app:adaptive} gives the decomposition and the spread over keys. A leaked key and a public seed both allow an exact simulation, so the leaked rows also stand for unkeyed sampling such as Nexus Sampling. Firstly, much of the damage of the answer attack needs no eviction. It lowers full memory from 86.7\% to 64.4\% and moves the Res of the earlier clips by at most 1.0\% under every quota. The excess over full memory lies above zero only for the base value policy, cap only with attention and KeyMem with a leaked key, and every interval of a secret key includes zero. An attack on eviction alone is weaker, and the token attack costs every quota at most 5.5\% and the base value policy 16.6\%.

Secondly, the secrecy of a key helps against the answer attack. Over the nine key pairs, a secret key keeps 8.4\% [4.3, 13.2], 9.5\% [5.2, 14.1] and 4.4\% [1.4, 7.8] more Early Acc than a leaked one for KeyMem with attention, value and merge, and 2.2\% [-0.1, 5.3], 4.3\% [0.4, 9.0] and 3.5\% [0.5, 6.7] for decide-once. Figure~\ref{fig:adaptive} shows that this gap of KeyMem with attention widens from 0.2\% with one choice to 8.4\% with 32 and 13.3\% [4.4, 24.4] with 64, while every curve keeps falling, so a secret key slows the damage over this range of search and does not stop it. Thirdly, a secret key does not make a rule stronger than the deterministic quotas. KeyMem trails equal spans by 19.4\% [12.8, 26.2] with attention and differs from cap only by 0.6\% [-4.3, 5.9] and -0.4\% [-4.6, 3.8] with value and merge. Decide-once with a secret key leads equal spans by 8.4\% [1.9, 15.3] with value and trails it by 14.4\% [8.3, 21.0] with attention, so the strongest rule depends on the base policy, and the order holds at $B=1024$ (Table~\ref{tab:b1024} in Appendix~\ref{app:adaptive}).

\begin{table}[t]
\centering
\footnotesize
\caption{Accuracy on the 200 s text streams by evidence age (s).}
\label{tab:long}
\setlength{\tabcolsep}{1.4pt}
\begin{tabular}{@{}lccccccccc@{}}
\toprule
\multirow{2}{*}{Rule} & \multicolumn{3}{c}{Attention} & \multicolumn{3}{c}{Value} & \multicolumn{3}{c}{Merge} \\
\cmidrule(lr){2-4}\cmidrule(lr){5-7}\cmidrule(lr){8-10}
 & All & 0--40 & ${>}$100 & All & 0--40 & ${>}$100 & All & 0--40 & ${>}$100 \\
\midrule
Full memory & 74.1 & 81.7 & 73.3 & 74.1 & 81.7 & 73.3 & 74.1 & 81.7 & 73.3 \\
\midrule
Base policy & 25.9 & 46.7 & 20.8 & 43.0 & 36.7 & 51.7 & 52.6 & 63.3 & 55.0 \\
Equal spans & 46.3 & 40.0 & \textbf{63.3} & 55.6 & 60.0 & 57.5 & 64.8 & 60.0 & \textbf{70.8} \\
Cap only & 48.1 & \textbf{80.0} & 34.2 & 55.6 & 56.7 & 56.7 & 57.8 & 66.7 & 59.2 \\
KeyMem & 48.1 & 78.3 & 35.8 & 46.3 & \textbf{78.3} & 35.0 & 57.8 & \textbf{80.0} & 53.3 \\
Decide-once & \textbf{58.1} & 73.3 & 55.0 & \textbf{61.9} & 73.3 & \textbf{58.3} & \textbf{65.2} & 70.0 & 65.8 \\
\bottomrule
\end{tabular}
\end{table}

\subsection{Long Streams}
\label{sec:long}

To test retention over many rounds, we extend the coverage streams to 18 clips, so that every stream lasts 200 s and the budget of $B=512$ holds 4\% of it, and we ask about every clip on 15 test streams of Qwen2.5-VL-7B with the calm or the text segment after the third clip. Table~\ref{tab:long} reports the text streams by the age of the evidence when the questions are asked, and Appendix~\ref{app:long} adds the middle ages, the Res, the calm streams, Qwen2.5-VL-3B at 200 s and Qwen2.5-VL-7B at 400 s. Firstly, KeyMem loses old evidence. After more than 25 rounds of fresh draws, KeyMem answers 27.5\% [15.0, 38.3], 22.5\% [13.3, 31.7] and 17.5\% [9.2, 26.7] fewer questions about clips older than 100 s than equal spans with attention, value and merge. Secondly, decide-once removes most of this erosion. It gains 19.2\% [10.0, 28.3], 23.3\% [11.7, 34.2] and 12.5\% [5.8, 20.0] over KeyMem on the oldest clips, and its comparison with equal spans there stays unresolved. The erosion repeats on Qwen2.5-VL-3B, with every interval below zero, and at 400 s with value and merge.

Long segments reach the limit of the cap. A 60 s text segment meets seven buckets, so Proposition 1 allows 672 tokens, more than the budget, and cap only with value falls to 24.1\%, while equal spans holds 52.2\% (Table~\ref{tab:long60} in Appendix~\ref{app:long}). An overlay on the whole stream defeats every temporal rule, since it spreads the failure over all of time.

\subsection{Choosing a Temporal Rule}
\label{sec:decision}

Table~\ref{tab:decision} in Appendix~\ref{app:decision} summarizes the four rules. Equal spans is a strong default in our tests, since it keeps a floor for every span, needs no key and keeps the most evidence older than 100 s. A hard cap is preferable when the share of a burst must be bounded from its first chunk, as when the stream length is unknown. A secret key is worth adding only when an adversary may simulate the memory, and then KeyMem suits a deployment that asks about the recent past, and decide-once suits one that also needs old evidence.

\section{Conclusion}
\label{sec:conclusion}

Score-based retention gives the content of a stream implicit control over where a finite memory is spent. Temporal occupancy control removes this failure, and both equal spans and the hard cap of KeyMem eliminate the observed concentration failures and exceed the worst-case accuracy of InfiniPot-V and Nexus Sampling. The hard cap adds a content-independent bound on the share of any segment from its first chunk but keeps no more earlier evidence than equal spans, which remains the accuracy-oriented default in our tests. A secret key makes retention less predictable to an adversary without uniformly improving accuracy over deterministic quotas, and decide-once sampling avoids much of the erosion caused by repeated draws. Robust streaming memory therefore begins with explicit temporal allocation, while secrecy remains an optional defense against an adversary who simulates the memory.

\noindent\textbf{Limitations.} Coverage uses one action question per UCF101 clip, and StreamingBench covers only Qwen2.5-VL-7B at one budget, and other model families and OVO-Bench~\cite{niu2025ovobench} remain untested. The adaptive study covers one model, and the published rules keep one retention set for all layers of the language model.

{
    \small
    \bibliographystyle{ieeenat_fullname}
    \bibliography{main}
}

\clearpage
\maketitlesupplementary
\nobalance
\appendix
\setcounter{topnumber}{2}
\setcounter{dbltopnumber}{1}
\setcounter{totalnumber}{3}
\renewcommand{\topfraction}{0.85}
\renewcommand{\dbltopfraction}{0.7}
\renewcommand{\textfraction}{0.2}
\raggedbottom

\section{Proofs}
\label{sec:proofs}

\noindent\textbf{Proof of Proposition 1.} Fix a segment $[t_0,t_0+L)$ of the stream and an aligned bucket length $W$. The buckets are the intervals $[kW,(k+1)W)$ for integers $k$, so the segment meets the buckets with indices from $\lfloor t_0/W\rfloor$ to $\lfloor(t_0+L)/W\rfloor$, and there are at most $\lceil L/W\rceil+1$ of them. After every chunk, the cap leaves at most $\lfloor\rho B\rfloor$ eligible tokens in each bucket, and the sampling step only removes tokens from the eligible set. Removed tokens never return, and a bucket gains tokens only when a chunk brings new frames from its interval, after which the cap is applied again. Every bucket therefore holds at most $\lfloor\rho B\rfloor$ tokens after every chunk, and summing over the buckets that the segment meets gives the bound. The argument uses neither the scores nor the random draws, so the bound holds for every content and every key. \hfill$\square$

\noindent\textbf{Proof of Proposition 2.} Fix round $t$, the eligible set $\mathcal{E}_t$ with $n$ tokens and the number of survivors $m$. The Gumbel-top-$k$ selection has the same distribution as $m$ sequential draws without replacement, where each draw picks a remaining token with probability proportional to its weight $w_j=e^{u_j/\tau}$~\cite{kool2019stochastic}. Because $u_j\in[0,1]$, every weight satisfies $1\le w_j\le a$ with $a=e^{1/\tau}$. Consider the $k$-th draw with token $i$ not yet picked. The pool then holds $i$ and $n-k$ other tokens, so the draw picks $i$ with probability
\begin{equation}
\frac{w_i}{w_i+\sum_{j\ne i}w_j}\ge\frac{1}{1+(n-k)a},
\end{equation}
since the left side grows with $w_i$ and falls with every $w_j$. Token $i$ is evicted only if none of the $m$ draws picks it, and the chain rule over the draws gives
\begin{equation}
\Pr[i\notin\mathcal{M}_t]\le\prod_{k=1}^{m}\frac{(n-k)a}{1+(n-k)a}=\prod_{j=r}^{n-1}\frac{ja}{1+ja},
\end{equation}
with $j=n-k$ and $r=n-m$. For the closed form, we use $\log(1-x)\le-x$ on every factor, which gives
\begin{align}
\log\Pr[i\notin\mathcal{M}_t]&\le-\sum_{j=r}^{n-1}\frac{1}{1+ja}\le-\int_r^n\frac{\mathrm{d}x}{1+ax}\nonumber\\
&=-\frac{1}{a}\log\frac{1+an}{1+ar},
\end{align}
because $1/(1+ax)$ is decreasing and the sum is a left Riemann sum of the integral. Exponentiating both sides yields the closed form of Equation~\ref{eq:retention} in the main text. \hfill$\square$

\noindent\textbf{Proof of Proposition 3.} Fix the ranks and a token $i$ that the cap keeps eligible through round $t$. The ranks and the perturbations do not change, so each token holds one value $z_j$ in every round. In round $s\le t$, the keyed step removes token $i$ only if $m_s=B$ tokens of $\mathcal{E}_s$ hold a larger value, since a round with at most $B$ eligible tokens removes none. Because $\mathcal{E}_s\subseteq\mathcal{A}_t$, token $i$ survives through round $t$ whenever fewer than $B$ tokens of $\mathcal{A}_t$ hold a larger value, which is the event that $i$ ranks among the top $\min(B,N)$ of $\mathcal{A}_t$. With fixed ranks, the cap depends on the ranks alone, so $\mathcal{A}_t$ does not depend on the perturbations. The perturbations of $\mathcal{A}_t$ are independent Gumbel variables when the hash acts as a random function, so the top $\min(B,N)$ form the Plackett-Luce draw of the proof of Proposition 2 with $n=N$ and $m=\min(B,N)$, and that proof gives the bound. \hfill$\square$

\noindent\textbf{Eligibility.} A token is ineligible only when its own bucket holds more than $\lfloor\rho B\rfloor$ tokens and it ranks below the cap within that bucket, so the cap decides deterministically within an overfull bucket, and the keyed draw decides among the buckets. A bucket receives new tokens only while the stream is inside its interval, so once the stream has moved on, its count can only fall, and every token that is still in it stays eligible in every later round. A segment that floods its own buckets therefore loses its lowest-ranked tokens for certain, while every older bucket keeps its eligible tokens and their per-round guarantee. Proposition 1 counts the aligned buckets that a segment meets, so a segment of length $W$ that straddles two buckets can hold up to $2\lfloor\rho B\rfloor$ tokens, and a tighter limit on every sliding window would need an interval-packing selector. All probabilities are taken over the uniform variables $\xi_i$, which treats the keyed generator as a random function of the key, and our implementation clamps every $\xi_i$ to $[10^{-7},1-10^{-7}]$, which changes the probability of any selection by less than $2\times10^{-4}$ at $n\le768$.

\section{Hyperparameter Selection and Harness}
\label{app:validation}
\suppressfloats[t]

\begin{table}[t]
\centering
\small
\caption{Validation accuracy of KeyMem.}
\label{tab:val}
\begin{tabular}{@{}ccccc@{}}
\toprule
$\tau$ & $\rho$ & Qwen2.5-VL-3B & Qwen2.5-VL-7B & Mean \\
\midrule
0.1 & 0.25 & 76.4 & 80.9 & 78.6 \\
0.25 & 0.25 & 77.2 & 82.1 & 79.6 \\
0.5 & 0.25 & 77.0 & 82.7 & \textbf{79.9} \\
0.75 & 0.25 & 75.5 & 81.9 & 78.7 \\
1.5 & 0.25 & 75.5 & 81.9 & 78.7 \\
\midrule
0.5 & 0.0625 & 74.2 & 79.9 & 77.1 \\
0.5 & 0.125 & 77.0 & 83.3 & 80.2 \\
0.5 & 0.1875 & 77.8 & 84.4 & \textbf{81.1} \\
0.5 & 0.25 & 77.0 & 82.7 & 79.9 \\
\bottomrule
\end{tabular}
\end{table}

Table~\ref{tab:val} reports the accuracy of KeyMem on the 12 validation streams, averaged over both Qwen models, the three base policies and the calm, text and cuts segments at the smaller budget. We first vary $\tau$ among 0.1, 0.25, 0.5, 0.75 and 1.5 with $\rho=0.25$ and select $\tau=0.5$, and we then vary $\rho$ among 0.0625, 0.125, 0.1875 and 0.25 with $\tau=0.5$ and select $\rho=0.1875$. The setting with $\tau=0.5$ and $\rho=0.25$ belongs to both searches, so it appears twice. The validation streams are built in the same way from the training split of UCF101, so the selection never sees a test clip, although the text and cuts stressors appear in both splits. The choice depends little on which stressors validate it. Leaving the text segment out keeps $\rho=0.1875$ and would prefer $\tau=0.25$ to $\tau=0.5$ by 0.3\%. We tuned nothing else, and $W$ and $R$ keep the values that we fixed before any experiment. To check our harness, we compare it with an unlimited budget against one offline forward pass over the whole prompt. On 50 Qwen2.5-VL-7B streams, the positions match exactly, the predicted answers agree on every stream, and the mean total variation between the two answer distributions is 0.0015. On 10 LLaVA-OneVision streams, the answers agree everywhere, with a mean total variation of 0.0016.

\section{First Adversary}
\label{app:first}

\begin{table}[t]
\centering
\small
\caption{Early accuracy and residency of the three earlier clips under the adversarial segment.}
\label{tab:adv}
\setlength{\tabcolsep}{1.5pt}
\begin{tabular}{@{}lcccccc@{}}
\toprule
\multirow{3}{*}{Policy} & \multicolumn{3}{c}{Qwen2.5-VL-3B} & \multicolumn{3}{c}{Qwen2.5-VL-7B} \\
\cmidrule(lr){2-4}\cmidrule(lr){5-7}
 & Calm & Adversary & \stk{Leaked\\key} & Calm & Adversary & \stk{Leaked\\key} \\
\midrule
\multicolumn{7}{@{}l}{Early Acc} \\
Attention & 67.8 & 66.7 & -- & 27.8 & 24.4 & -- \\
\quad + KeyMem & 77.8 & 76.7 & 76.7 & 72.2 & 80.0 & 77.8 \\
Value & 56.7 & 48.9 & -- & 73.3 & 57.8 & -- \\
\quad + KeyMem & 52.2 & 51.1 & 51.1 & 76.7 & 73.3 & 73.3 \\
Merge & 74.4 & 71.1 & -- & 87.8 & 77.8 & -- \\
\quad + KeyMem & 71.1 & 70.0 & 71.1 & 83.3 & 80.0 & 81.1 \\
\midrule
\multicolumn{7}{@{}l}{Res} \\
Attention & 1.7 & 1.4 & -- & 0.2 & 0.1 & -- \\
\quad + KeyMem & 6.9 & 6.7 & -- & 6.4 & 6.1 & -- \\
Value & 9.5 & 8.4 & -- & 9.0 & 1.9 & -- \\
\quad + KeyMem & 7.5 & 7.1 & -- & 7.7 & 6.2 & -- \\
Merge & 11.7 & 10.7 & -- & 13.4 & 10.2 & -- \\
\quad + KeyMem & 8.4 & 8.3 & -- & 8.8 & 8.4 & -- \\
\bottomrule
\end{tabular}
\end{table}

The first adversary inserts each of the 16 candidates after the third clip, keeps the candidate that leaves the fewest tokens of the three earlier clips in memory, and uses a key of its own against KeyMem. Table~\ref{tab:adv} reports the Early Acc of each policy at the 8 s budget on all 30 test streams, with the calm control and with the selected segment. Against deterministic retention, the adversary lowers the Early Acc by 6.9\% on average over the two models and three base policies, and by 15.5\% for the value policy on Qwen2.5-VL-7B, where it picks a text segment in all 30 streams and leaves the earlier clips with 1.9\% of their tokens instead of 9.0\%. Merge on Qwen2.5-VL-7B loses 10.0\%, which shows that the most robust policy under average conditions is not safe against a chosen segment. With KeyMem, no Early Acc falls by more than 3.4\%, the six changes average to a loss of 0.4\%, and the earlier clips of the value policy on Qwen2.5-VL-7B keep 6.2\% of their tokens instead of 7.7\%. Two cells of KeyMem with attention rise under the adversary, which shows that the count of surviving tokens is only a proxy for accuracy. The hard cap bounds what any segment can take whatever the adversary predicts, so against this token-counting adversary, the protection does not rest on the secrecy of the key. Under the adversary, KeyMem leads its base policy in five of the six cells, although merge on both models and value on Qwen2.5-VL-3B start higher.

The lower half of Table~\ref{tab:adv} reports the Res of the three clips before the adversarial segment for the same runs. The adversary cuts the residency of the deterministic value policy on Qwen2.5-VL-7B from 9.0\% to 1.9\%, while no policy with KeyMem loses more than 1.5\%. The leaked-key columns of Table~\ref{tab:adv} repeat the adversarial runs with an adversary that holds the defender's key, which changes no Early Acc of KeyMem by more than 2.2\%.

\section{Shifted Grid and Long Segments}
\label{app:extra}

\begin{table}[t]
\centering
\small
\caption{Accuracy with a 40 s segment and with a shifted grid.}
\label{tab:extra}
\begin{tabular}{@{}lccc@{}}
\toprule
Policy & Calm & Cuts & Text \\
\midrule
\multicolumn{4}{@{}l}{40 s segment} \\
\quad Attention & 44.4 & 42.2 & 42.2 \\
\quad Attention + KeyMem & 73.3 & 74.4 & 73.3 \\
\quad Value & 78.9 & 66.7 & 37.8 \\
\quad Value + KeyMem & 73.3 & 73.3 & 71.1 \\
\quad Merge & 81.1 & 82.2 & 75.6 \\
\quad Merge + KeyMem & 83.3 & 85.6 & 80.0 \\
\midrule
\multicolumn{4}{@{}l}{Grid shifted by 5 s} \\
\quad Attention + KeyMem & 86.7 & 83.3 & 82.2 \\
\quad Value & 85.6 & 77.8 & 60.0 \\
\quad Value + KeyMem & 86.7 & 82.2 & 76.7 \\
\quad Merge + KeyMem & 84.4 & 87.8 & 87.8 \\
\bottomrule
\end{tabular}
\end{table}
\begin{table*}[t]
\centering
\footnotesize
\caption{Accuracy and early accuracy in calm streams and at worst over the five stressors.}
\label{tab:main}
\setlength{\tabcolsep}{2.6pt}
\begin{tabular}{@{}llccccccccccccc@{}}
\toprule
\multirow{3}{*}{Budget} & \multirow{3}{*}{Policy} & \multicolumn{7}{c}{Acc} & \multicolumn{6}{c}{Early Acc} \\
\cmidrule(lr){3-9}\cmidrule(lr){10-15}
 & & \multicolumn{2}{c}{Qwen2.5-VL-3B} & \multicolumn{2}{c}{Qwen2.5-VL-7B} & \multicolumn{2}{c}{LLaVA-OneVision-7B} & \multirow{2}{*}{All} & \multicolumn{2}{c}{Qwen2.5-VL-3B} & \multicolumn{2}{c}{Qwen2.5-VL-7B} & \multicolumn{2}{c}{LLaVA-OneVision-7B} \\
\cmidrule(lr){3-4}\cmidrule(lr){5-6}\cmidrule(lr){7-8}\cmidrule(lr){10-11}\cmidrule(lr){12-13}\cmidrule(lr){14-15}
 & & Calm & Worst & Calm & Worst & Calm & Worst & & Calm & Worst & Calm & Worst & Calm & Worst \\
\midrule
\multirow{9}{*}{8 s} & Sliding window & 32.2 & 26.7 & 40.0 & 34.4 & 36.1 & 32.8 & 26.7 & 25.6 & 12.2 & 28.9 & 20.0 & 24.4 & 16.7 \\
 & Uniform & 58.9 & 48.9 & 51.7 & 50.6 & 58.9 & 53.3 & 48.9 & 33.3 & 18.9 & 25.6 & 23.3 & 31.1 & 24.4 \\
 & Attention & 75.0 & 70.6 & 45.6 & 42.2 & 50.6 & 43.3 & 42.2 & 67.8 & 64.4 & 27.8 & 22.2 & 33.3 & 32.2 \\
 & \quad + KeyMem & \textbf{81.7} & \textbf{76.1} & 80.0 & 79.4 & \textbf{77.2} & \textbf{75.6} & \textbf{75.6} & \textbf{77.8} & \textbf{71.1} & 72.2 & 71.1 & 62.2 & 61.1 \\
 & Value & 57.2 & 53.3 & 79.4 & 58.3 & 72.2 & 65.0 & 53.3 & 56.7 & 51.1 & 73.3 & 58.9 & 61.1 & 51.1 \\
 & \quad + KeyMem & 58.3 & 52.8 & 82.8 & 77.2 & 73.9 & 71.1 & 52.8 & 52.2 & 43.3 & 76.7 & 72.2 & 58.9 & 56.7 \\
 & Merge & 73.3 & 69.4 & 86.1 & 79.4 & 76.7 & 75.0 & 69.4 & 74.4 & 70.0 & \textbf{87.8} & \textbf{85.6} & 62.2 & \textbf{62.2} \\
 & \quad + KeyMem & 75.6 & 72.8 & \textbf{86.7} & \textbf{82.2} & 75.6 & 70.6 & 70.6 & 71.1 & 66.7 & 83.3 & 80.0 & \textbf{63.3} & 55.6 \\
 & Full memory & 84.4 & 80.6 & 90.6 & 83.3 & 86.7 & 80.0 & 80.0 & 82.2 & 80.0 & 86.7 & 81.1 & 80.0 & 74.4 \\
\midrule
\multirow{8}{*}{16 s} & Sliding window & 48.3 & 42.8 & 48.9 & 43.9 & 52.8 & 48.3 & 42.8 & 31.1 & 18.9 & 27.8 & 21.1 & 32.2 & 25.6 \\
 & Uniform & 68.9 & 63.9 & 70.6 & 66.1 & 69.4 & 62.8 & 62.8 & 46.7 & 37.8 & 47.8 & 45.6 & 45.6 & 38.9 \\
 & Attention & 82.8 & 75.0 & 66.7 & 60.0 & 78.3 & 69.4 & 60.0 & 81.1 & 73.3 & 46.7 & 37.8 & \textbf{67.8} & 60.0 \\
 & \quad + KeyMem & \textbf{86.1} & \textbf{80.6} & 82.2 & 79.4 & 79.4 & \textbf{77.8} & \textbf{77.8} & \textbf{84.4} & \textbf{82.2} & 71.1 & 72.2 & \textbf{67.8} & \textbf{65.6} \\
 & Value & 65.0 & 62.8 & 82.8 & 80.0 & 73.9 & 75.0 & 62.8 & 62.2 & 57.8 & 77.8 & 76.7 & 58.9 & 61.1 \\
 & \quad + KeyMem & 71.1 & 67.2 & 85.0 & 81.1 & 75.0 & 74.4 & 67.2 & 65.6 & 58.9 & 76.7 & 72.2 & 60.0 & 60.0 \\
 & Merge & 80.0 & 72.8 & 85.0 & \textbf{82.8} & 78.9 & \textbf{77.8} & 72.8 & 82.2 & 73.3 & \textbf{83.3} & \textbf{83.3} & 65.6 & 64.4 \\
 & \quad + KeyMem & 81.1 & 74.4 & \textbf{87.2} & 82.2 & \textbf{80.0} & 77.2 & 74.4 & 76.7 & 68.9 & 82.2 & 82.2 & 66.7 & \textbf{65.6} \\
\bottomrule
\end{tabular}
\end{table*}

We also run a targeted adversary that picks, among the same 16 candidates, the segment that leaves the fewest tokens of the first clip alone. On 12 streams, it keeps the accuracy of KeyMem with value on that clip at 75.0\% and lowers attention from 83.3\% to 75.0\%, which is one question in twelve, and it lowers neither policy behind the cap-only variant or when the adversary holds the leaked key.

Table~\ref{tab:extra} reports two stress tests of the cap on the first 15 test streams of Qwen2.5-VL-7B at $B=512$. The 40 s segment doubles the stress segment after the third clip, so every stream lasts 100 s and the segment fills four buckets of the cap. The shifted grid moves every bucket boundary of the cap by 5 s, so the 20 s segment straddles three buckets instead of filling two. The value policy alone uses no grid, so its row repeats the aligned run on the same streams. On the same 15 streams with the aligned grid, KeyMem reaches 83.3\%, 84.4\% and 88.9\% in calm streams and 80.0\%, 86.7\% and 83.3\% on the text segment with attention, value and merge.

\section{Decide-Once on Natural Stress}
\label{app:stress}

Figure~\ref{fig:stress} in the main text plots the accuracy of the three score-based policies with and without KeyMem on every model and stress type at the 8 s budget. Decide-once reaches a Worst Acc of 70.0\% on Qwen2.5-VL-3B, 80.9\% on Qwen2.5-VL-7B and 73.9\% on LLaVA-OneVision-7B, averaged over the three base policies (Table~\ref{tab:sota} in the main text). Its Worst Acc leads cap only by 6.1\% [1.1, 12.2] with merge on Qwen2.5-VL-3B and by 4.4\% [0.6, 9.4] with value on Qwen2.5-VL-7B, and the other 25 intervals against equal spans, cap only and KeyMem over the three models include zero, so these runs do not separate it from the other quotas on natural stress.

\section{Results at Both Budgets}
\label{app:full}

Table~\ref{tab:main} reports every policy with the calm control and its Worst Acc over the five stressors, and the column All takes the worst case over the three models as well. Over the 18 combinations of model, base policy and budget, KeyMem raises the Worst Acc in 13, by up to 37.2\%, with 7 paired intervals above zero, and it raises the calm accuracy in 17. At the 8 s budget, KeyMem raises this worst case from 42.2\% to 75.6\% for attention, which makes attention with KeyMem the most robust of these policies, and at the 16 s budget, it raises the worst case of all three base policies.

\section{Early Accuracy}
\label{app:early}

The right half of Table~\ref{tab:main} repeats the left half for the three clips that precede the stress segment. These clips are the oldest when the questions are asked, so they show most clearly how a policy treats old evidence. KeyMem raises the Early Acc of attention on both 7B models, and the gain reaches 44.4\% in calm streams on Qwen2.5-VL-7B at the 8 s budget. It lowers the Early Acc of value and merge in 15 of their 24 cells, by up to 7.8\% for value on Qwen2.5-VL-3B. A deterministic top-$k$ rule keeps the same high-scoring old tokens in every round, while a keyed draw evicts each of them with a small probability in every round, and these evictions compound over the dozen rounds that follow the earlier clips. A smaller $\tau$ reduces the compounding but also weakens the guarantee of Proposition 2, and Appendix~\ref{app:validation} selects $\tau=0.5$ by the accuracy over all clips.

\section{Statistical Uncertainty}
\label{app:boot}

\begin{table}[t]
\centering
\caption{Change in worst accuracy with KeyMem against each rule and its 95\% interval.}
\label{tab:boot}
\resizebox{\columnwidth}{!}{%
\begin{tabular}{@{}lcccc@{}}
\toprule
\multirow{2}{*}{Policy} & \multicolumn{2}{c}{Against the base policy} & Against equal spans & Against cap only \\
\cmidrule(lr){2-3}\cmidrule(lr){4-4}\cmidrule(lr){5-5}
 & 8 s & 16 s & 8 s & 8 s \\
\midrule
\multicolumn{5}{@{}l}{Qwen2.5-VL-3B} \\
\quad Attention & 5.6 [1.1, 11.1] & 5.6 [0.6, 10.6] & 2.8 [-1.7, 6.7] & -1.7 [-3.3, 1.1] \\
\quad Value & -0.6 [-7.8, 8.9] & 4.4 [0.0, 9.4] & -0.6 [-7.2, 7.2] & -2.8 [-7.8, 4.4] \\
\quad Merge & 3.3 [-2.2, 7.8] & 1.7 [-3.3, 6.7] & -0.6 [-6.1, 3.9] & 4.4 [-1.1, 8.3] \\
\midrule
\multicolumn{5}{@{}l}{Qwen2.5-VL-7B} \\
\quad Attention & 37.2 [30.0, 43.3] & 19.4 [13.9, 25.6] & -0.6 [-7.2, 5.6] & 1.1 [-3.9, 5.0] \\
\quad Value & 18.9 [11.1, 23.9] & 1.1 [-2.2, 5.6] & -2.2 [-6.1, 1.1] & 2.2 [-1.7, 5.6] \\
\quad Merge & 2.8 [-2.2, 5.6] & -0.6 [-3.3, 5.0] & 0.6 [-3.9, 2.8] & -1.7 [-4.4, 0.6] \\
\midrule
\multicolumn{5}{@{}l}{LLaVA-OneVision-7B} \\
\quad Attention & 32.2 [23.9, 38.3] & 8.3 [0.6, 12.2] & 1.1 [-3.9, 6.1] & -0.6 [-3.3, 1.7] \\
\quad Value & 6.1 [-0.6, 11.7] & -0.6 [-3.3, 2.8] & 1.1 [-2.8, 3.9] & 0.0 [-2.8, 3.3] \\
\quad Merge & -4.4 [-7.2, 0.6] & -0.6 [-3.3, 3.3] & -3.9 [-7.2, 1.1] & -3.9 [-7.2, 1.1] \\
\bottomrule
\end{tabular}}
\end{table}

Table~\ref{tab:boot} gives paired bootstrap intervals over the 30 test streams, with 10,000 resamples, for the change in Worst Acc that KeyMem brings against the base policy at both budgets and against the two deterministic quotas at the 8 s budget. Each resample draws the streams with replacement and recomputes the Worst Acc of both policies as the minimum over the five stressors, so the intervals include the choice of the worst stressor. Against the base policy, 7 of the 18 intervals lie above zero, and all of them belong to attention or to value. No interval lies below zero, although the interval of value on Qwen2.5-VL-3B at the 16 s budget starts at zero. Against equal spans and against cap only, every one of the 18 intervals contains zero, and the calm accuracy behaves in the same way, with every interval against either quota between -8.3\% and 7.2\%.

\begin{table}[t]
\centering
\small
\caption{Accuracy of KeyMem with three secret keys.}
\label{tab:keys}
\begin{tabular}{@{}lcccccc@{}}
\toprule
\multirow{2}{*}{Policy} & \multicolumn{3}{c}{Calm} & \multicolumn{3}{c}{Worst Acc} \\
\cmidrule(lr){2-4}\cmidrule(lr){5-7}
 & Key 1 & Key 2 & Key 3 & Key 1 & Key 2 & Key 3 \\
\midrule
Attention & 80.0 & 78.3 & 78.9 & 79.4 & 80.6 & 81.1 \\
Value & 82.8 & 82.2 & 78.9 & 77.2 & 79.4 & 77.8 \\
Merge & 86.7 & 86.1 & 85.6 & 82.2 & 82.2 & 82.2 \\
\bottomrule
\end{tabular}
\end{table}

Table~\ref{tab:keys} repeats the 8 s runs of KeyMem on Qwen2.5-VL-7B with two further secret keys. The three keys change no Calm accuracy or Worst Acc by more than 3.9\%, which is smaller than most intervals over streams in Table~\ref{tab:boot}.

\section{Ablation}
\label{app:ablation}

\begin{table}[t]
\centering
\small
\caption{Ablation of KeyMem on Qwen2.5-VL-7B at $B=512$.}
\label{tab:ablation}
\setlength{\tabcolsep}{2.4pt}
\begin{tabular}{@{}lcccccc@{}}
\toprule
\multirow{3}{*}{Variant} & \multicolumn{3}{c}{Attention} & \multicolumn{3}{c}{Value} \\
\cmidrule(lr){2-4}\cmidrule(lr){5-7}
 & Calm & Adversary & \stk{Leaked\\key} & Calm & Adversary & \stk{Leaked\\key} \\
\midrule
KeyMem & 77.8 & 84.4 & 82.2 & 80.0 & 75.6 & 75.6 \\
Cap only & 71.1 & 75.6 & 75.6 & 77.8 & 84.4 & 80.0 \\
Key only & 26.7 & 28.9 & 26.7 & 57.8 & 42.2 & 42.2 \\
\bottomrule
\end{tabular}
\end{table}

Table~\ref{tab:ablation} removes one component of KeyMem at a time and reports the Early Acc on the first 15 test streams of Qwen2.5-VL-7B at $B=512$. The cap-only variant sets $\tau=0.01$, so the draw follows the ranks almost deterministically, and the key-only variant sets $\rho=1$, which removes the cap. The adversary is the first adversary of Appendix~\ref{app:first}. Without the cap, the keyed draw does not protect old evidence. Attention then keeps no token of the earlier clips even in calm streams, and the adversary lowers value from 57.8\% to 42.2\%. The cap alone recovers the protection, and the adversary lowers neither policy behind it. The keyed draw on top of the cap adds 6.7\% for attention and 2.2\% for value in calm streams, and under the adversary, the two variants trade places, which 15 streams cannot resolve. Its leaked-key columns repeat the adversarial runs with an adversary that holds the defender's key, which changes the Early Acc of either variant by at most 4.4\%.

\section{Burst Placement}
\label{app:burst}

\begin{figure}[t]
\centering
\includegraphics[width=\columnwidth]{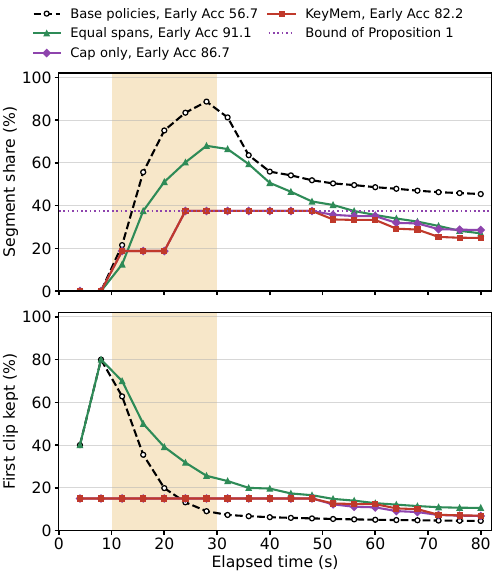}
\caption{Memory share of the text segment and tokens of the first clip kept after every chunk on Qwen2.5-VL-7B at $B=512$, averaged over the three base policies and the 30 test streams.}
\label{fig:burst}
\end{figure}

This section moves the 20 s text segment through the 80 s streams of Qwen2.5-VL-7B at $B=512$ on the 30 test streams. The segment follows the first, second, third or fifth clip, a bucket grid shifted by 5 s makes it meet three buckets, and a 40 s segment follows the first clip. The bound is the one of Proposition 1 for the buckets that the segment meets. Equal spans and the base policies use no bucket grid, so the shifted rows of Table~\ref{tab:burst} in the main text repeat their aligned values. KeyMem and decide-once reach the same peak share as cap only, except after the fifth clip, where they reach 35.9\% and 34.6\%. Over the seven placements of the burst in the order of that table, full memory reaches an Early Acc of 93.3\%, 86.7\%, 86.7\%, 84.0\%, 93.3\%, 86.7\% and 93.3\%, and the base policies reach 56.7\%, 57.2\%, 58.1\%, 53.1\%, 56.7\%, 57.2\% and 47.8\%.

\begin{table}[t]
\centering
\footnotesize
\caption{Accuracy of the published rules.}
\label{tab:sotafull}
\begin{tabular}{@{}lcccc@{}}
\toprule
\multirow{2}{*}{Rule} & \multicolumn{2}{c}{8 s} & \multicolumn{2}{c}{16 s} \\
\cmidrule(lr){2-3}\cmidrule(lr){4-5}
& Calm & Worst Acc & Calm & Worst Acc \\
\midrule
\multicolumn{5}{@{}l}{Qwen2.5-VL-3B} \\
\quad InfiniPot-V & 70.6 & 60.0 & 77.8 & 70.0 \\
\quad Nexus Sampling & 37.2 & 32.8 & 56.7 & 51.1 \\
\multicolumn{5}{@{}l}{Qwen2.5-VL-7B} \\
\quad InfiniPot-V & 70.6 & 67.2 & 78.3 & 72.2 \\
\quad Nexus Sampling & 43.3 & 38.3 & 51.7 & 50.6 \\
\multicolumn{5}{@{}l}{LLaVA-OneVision-7B} \\
\quad InfiniPot-V & 70.6 & 63.9 & 76.7 & 72.8 \\
\quad Nexus Sampling & 46.7 & 43.3 & 70.0 & 63.3 \\
\bottomrule
\end{tabular}
\end{table}

\begin{table}[t]
\centering
\footnotesize
\caption{Paired differences between every quota and InfiniPot-V with 95\% intervals.}
\label{tab:sotaci}
\resizebox{\columnwidth}{!}{%
\begin{tabular}{@{}lccc@{}}
\toprule
Quota & Qwen2.5-VL-3B & Qwen2.5-VL-7B & LLaVA-OneVision-7B \\
\midrule
\multicolumn{4}{@{}l}{Worst Acc} \\
\quad Attention + equal spans & 13.3 [6.1, 21.1] & 12.8 [7.2, 19.4] & 10.6 [2.8, 16.7] \\
\quad Attention + cap only & 17.8 [10.6, 24.4] & 11.1 [6.1, 17.2] & 12.2 [4.4, 18.3] \\
\quad Attention + KeyMem & 16.1 [8.9, 23.9] & 12.2 [6.7, 18.3] & 11.7 [3.9, 17.8] \\
\quad Value + equal spans & -6.7 [-14.4, 0.6] & 12.2 [8.9, 18.3] & 6.1 [0.0, 13.3] \\
\quad Value + cap only & -4.4 [-12.8, 1.7] & 7.8 [4.4, 13.3] & 7.2 [0.0, 12.8] \\
\quad Value + KeyMem & -7.2 [-13.9, 0.0] & 10.0 [6.7, 15.0] & 7.2 [0.6, 13.9] \\
\quad Merge + equal spans & 13.3 [4.4, 20.0] & 14.4 [10.6, 21.1] & 10.6 [3.3, 15.6] \\
\quad Merge + cap only & 8.3 [1.1, 14.4] & 16.7 [12.2, 22.2] & 10.6 [2.8, 16.1] \\
\quad Merge + KeyMem & 12.8 [4.4, 18.3] & 15.0 [10.0, 20.6] & 6.7 [0.0, 12.8] \\
\midrule
\multicolumn{4}{@{}l}{Corruption} \\
\quad Attention + equal spans & 15.3 [10.8, 20.0] & 14.7 [8.6, 20.8] & 7.5 [1.1, 13.9] \\
\quad Attention + cap only & 15.0 [9.4, 21.1] & 10.6 [5.0, 16.4] & 6.4 [2.2, 10.8] \\
\quad Attention + KeyMem & 15.8 [10.3, 21.7] & 11.7 [5.8, 18.1] & 4.7 [0.3, 9.4] \\
\quad Value + equal spans & -6.9 [-14.4, 0.6] & 11.7 [6.4, 17.2] & 3.6 [-1.4, 8.9] \\
\quad Value + cap only & -8.6 [-15.3, -2.5] & 11.4 [6.4, 16.4] & -0.3 [-5.0, 4.2] \\
\quad Value + KeyMem & -3.9 [-9.7, 1.9] & 11.4 [6.1, 16.9] & 0.8 [-3.9, 5.8] \\
\quad Merge + equal spans & 8.6 [3.6, 13.9] & 13.6 [7.8, 19.7] & 6.1 [0.6, 11.7] \\
\quad Merge + cap only & 8.6 [2.8, 14.4] & 13.6 [8.1, 19.4] & 0.6 [-4.2, 5.6] \\
\quad Merge + KeyMem & 11.1 [5.6, 16.7] & 13.1 [7.5, 18.9] & 1.9 [-2.5, 6.4] \\
\bottomrule
\end{tabular}}
\end{table}

\begin{table}[t]
\centering
\footnotesize
\caption{Accuracy under the two corruptions.}
\label{tab:corrupt}
\resizebox{\columnwidth}{!}{%
\begin{tabular}{@{}lcccccc@{}}
\toprule
\multirow{2}{*}{Policy} & \multicolumn{2}{c}{Qwen2.5-VL-3B} & \multicolumn{2}{c}{Qwen2.5-VL-7B} & \multicolumn{2}{c}{LLaVA-OneVision-7B} \\
\cmidrule(lr){2-3}\cmidrule(lr){4-5}\cmidrule(lr){6-7}
& Noise & Blur & Noise & Blur & Noise & Blur \\
\midrule
Full memory & 84.4 & 84.4 & 89.4 & 87.8 & 87.8 & 86.1 \\
InfiniPot-V & 68.3 & 66.1 & 70.6 & 71.1 & 73.9 & 73.3 \\
Nexus Sampling & 36.7 & 37.2 & 42.8 & 42.8 & 45.0 & 47.2 \\
\midrule
Attention & 76.7 & 76.1 & 48.9 & 45.0 & 51.1 & 53.9 \\
\quad + Equal spans & 81.7 & 83.3 & 85.0 & 86.1 & 80.6 & 81.7 \\
\quad + Cap only & 81.7 & 82.8 & 81.7 & 81.1 & 78.9 & 81.1 \\
\quad + KeyMem & 81.7 & 84.4 & 82.2 & 82.8 & 77.8 & 78.9 \\
\quad + Decide-once & 81.7 & 83.9 & 82.2 & 81.1 & 78.3 & 78.9 \\
\midrule
Value & 57.8 & 56.1 & 79.4 & 83.3 & 73.3 & 74.4 \\
\quad + Equal spans & 62.8 & 57.8 & 82.2 & 82.8 & 76.1 & 78.3 \\
\quad + Cap only & 60.0 & 57.2 & 80.6 & 83.9 & 72.2 & 74.4 \\
\quad + KeyMem & 63.9 & 62.8 & 82.2 & 82.2 & 72.2 & 76.7 \\
\quad + Decide-once & 62.2 & 62.2 & 81.7 & 85.0 & 75.0 & 76.7 \\
\midrule
Merge & 71.1 & 73.9 & 85.6 & 86.7 & 77.8 & 77.2 \\
\quad + Equal spans & 75.0 & 76.7 & 84.4 & 84.4 & 78.9 & 80.6 \\
\quad + Cap only & 76.1 & 75.6 & 85.0 & 83.9 & 74.4 & 73.9 \\
\quad + KeyMem & 77.8 & 78.9 & 83.9 & 83.9 & 75.6 & 75.6 \\
\quad + Decide-once & 76.7 & 77.2 & 82.8 & 81.7 & 76.7 & 75.0 \\
\bottomrule
\end{tabular}}
\end{table}

The bound does not protect the earlier clips. Each earlier clip fills one aligned bucket of 640 tokens, and the cap leaves 96 of them, so the capped rules keep at most 15.0\% of an earlier clip, fill only 50.0\% to 77.3\% of the memory until the segment ends, and at its end keep 2.2\% to 8.2\% fewer tokens of the earlier clips than equal spans. With each stream averaged over the three base policies, cap only and KeyMem trail equal spans on Early Acc in all seven placements, by 3.3\% to 15.6\% and 1.1\% to 15.6\%, with 11 of their 14 paired intervals below zero, and decide-once differs by -7.8\% to 1.1\%, with 2 of its 7 intervals below zero. No capped rule leads equal spans with an interval above zero, and with the calm segment, cap only and KeyMem also trail equal spans at all four positions. The fixed bound thus brings no gain in retention or accuracy over equal spans.

Figure~\ref{fig:burst} follows the running example of the introduction in the main text after every chunk of 4 s, with the text segment from 10 s to 30 s shaded. Under the base policies and equal spans, the segment takes most of the memory while it plays, and equal spans still keeps the most tokens of the first clip. The cap holds the bucket of the first clip to 96 of its 640 tokens from the first chunk, while most of the memory is still empty, so the capped rules fill only 18.8\% to 75.0\% of the memory during the first 40 s. When the segment ends, equal spans keeps 23.2\% of the first clip, the capped rules 15.0\% and the base policies 7.3\%, and at the end of the stream, equal spans still keeps 10.6\%, against 6.8\% to 8.6\% for the capped rules and 4.5\% for the base policies.

\begin{table*}[t]
\centering
\footnotesize
\caption{Decomposition of the attacks with 32 choices on Qwen2.5-VL-7B at $B=512$.}
\label{tab:mech}
\setlength{\tabcolsep}{3pt}
\begin{tabular}{@{}lcccccccc@{}}
\toprule
\multirow{2}{*}{Defense} & \multicolumn{4}{c}{Answer attack} & \multicolumn{4}{c}{Token attack} \\
\cmidrule(lr){2-5}\cmidrule(lr){6-9}
 & Change in Res & Total drop & Full drop & Excess & Change in Res & Total drop & Full drop & Excess \\
\midrule
\multicolumn{9}{@{}l}{Attention} \\
\quad Base policy & 0.0 & 5.6 & 5.6 & 0.0 [-7.8, 7.8] & -0.1 & 2.2 & 7.8 & -5.6 [-13.3, 2.2] \\
\quad Equal spans & 0.3 & 10.0 & 13.3 & -3.3 [-10.0, 4.4] & -0.3 & 2.2 & 5.6 & -3.3 [-11.1, 3.3] \\
\quad Cap only & -1.0 & 23.3 & 11.1 & 12.2 [1.1, 23.3] & -1.7 & 4.4 & 3.3 & 1.1 [-10.0, 13.3] \\
\quad KeyMem, leaked key & -0.3 & 23.7 & 11.9 & 11.9 [3.7, 20.0] & -0.7 & -1.1 & 3.3 & -4.4 [-11.5, 3.0] \\
\quad KeyMem, secret key & -0.2 & 15.3 & 13.7 & 1.6 [-6.0, 9.0] & -0.6 & -1.1 & 1.1 & -2.2 [-9.5, 5.2] \\
\quad Decide-once, leaked key & -0.1 & 18.1 & 14.4 & 3.7 [-5.2, 11.9] & -0.5 & -2.2 & 7.0 & -9.3 [-18.1, -0.7] \\
\quad Decide-once, secret key & -0.1 & 15.9 & 13.7 & 2.2 [-5.2, 9.3] & -0.2 & -0.2 & 5.2 & -5.4 [-11.2, 0.2] \\
\midrule
\multicolumn{9}{@{}l}{Value} \\
\quad Base policy & -3.3 & 27.8 & 10.0 & 17.8 [6.7, 30.0] & -7.1 & 16.7 & 2.2 & 14.4 [0.0, 27.8] \\
\quad Equal spans & 0.0 & 21.1 & 12.2 & 8.9 [-1.1, 18.9] & -0.3 & 3.3 & 5.6 & -2.2 [-8.9, 3.3] \\
\quad Cap only & -0.1 & 16.7 & 13.3 & 3.3 [-5.6, 13.3] & -1.5 & 2.2 & 1.1 & 1.1 [-6.7, 8.9] \\
\quad KeyMem, leaked key & -0.2 & 23.0 & 12.6 & 10.4 [1.5, 19.3] & -2.0 & 0.4 & 0.0 & 0.4 [-6.3, 6.3] \\
\quad KeyMem, secret key & -0.1 & 13.5 & 13.3 & 0.1 [-7.5, 7.7] & -2.0 & 0.9 & 0.0 & 0.9 [-5.3, 6.7] \\
\quad Decide-once, leaked key & 0.0 & 16.7 & 13.7 & 3.0 [-3.3, 8.9] & -0.9 & -2.2 & 2.2 & -4.4 [-11.9, 2.2] \\
\quad Decide-once, secret key & 0.0 & 12.3 & 14.8 & -2.5 [-8.3, 2.8] & -0.8 & -3.6 & 0.7 & -4.3 [-11.7, 2.1] \\
\midrule
\multicolumn{9}{@{}l}{Merge} \\
\quad Base policy & -1.0 & 16.7 & 15.6 & 1.1 [-6.7, 8.9] & -3.7 & 7.8 & 0.0 & 7.8 [-1.1, 16.7] \\
\quad Equal spans & 0.0 & 13.3 & 16.7 & -3.3 [-11.1, 4.4] & -0.3 & 5.6 & 7.8 & -2.2 [-8.9, 4.4] \\
\quad Cap only & -0.1 & 15.6 & 16.7 & -1.1 [-10.0, 6.7] & -0.7 & 3.3 & 2.2 & 1.1 [-7.8, 10.0] \\
\quad KeyMem, leaked key & -0.1 & 15.6 & 14.8 & 0.7 [-6.7, 7.8] & -0.7 & 3.3 & 2.6 & 0.7 [-5.9, 7.0] \\
\quad KeyMem, secret key & -0.1 & 11.1 & 14.8 & -3.7 [-11.2, 3.5] & -0.5 & 1.0 & 1.5 & -0.5 [-6.5, 5.6] \\
\quad Decide-once, leaked key & -0.1 & 12.2 & 14.8 & -2.6 [-10.7, 5.6] & -0.5 & 3.0 & 4.4 & -1.5 [-7.4, 4.1] \\
\quad Decide-once, secret key & 0.0 & 8.8 & 13.3 & -4.6 [-11.5, 3.0] & -0.2 & 0.5 & 3.3 & -2.8 [-8.8, 3.6] \\
\bottomrule
\end{tabular}
\end{table*}

\begin{table*}[t]
\centering
\footnotesize
\caption{Secret and leaked keys over nine key pairs against the answer attack.}
\label{tab:pairs}
\setlength{\tabcolsep}{3pt}
\begin{tabular}{@{}lccccccccc@{}}
\toprule
Defense & Secret & Range over pairs & Leaked & Averaged & Gap & Range of gap & Streams and keys & Keys only & Gap, token attack \\
\midrule
KeyMem, attention & 56.2 & 51.1 to 58.9 & 47.8 & 54.8 & 8.4 & 4.4 to 11.1 & [4.3, 13.2] & [7.2, 9.5] & 0.0 [-4.6, 4.3] \\
KeyMem, value & 61.7 & 57.8 to 64.4 & 52.2 & 58.1 & 9.5 & 6.7 to 12.2 & [5.2, 14.1] & [8.4, 10.6] & -0.5 [-2.2, 1.0] \\
KeyMem, merge & 71.9 & 70.0 to 73.3 & 67.4 & 72.2 & 4.4 & 1.1 to 7.8 & [1.4, 7.8] & [3.1, 5.9] & 2.3 [-0.4, 5.6] \\
Decide-once, attention & 61.1 & 57.8 to 62.2 & 58.9 & 60.4 & 2.2 & 0.0 to 3.3 & [-0.1, 5.3] & [1.4, 3.0] & -2.0 [-6.8, 2.8] \\
Decide-once, value & 67.3 & 64.4 to 72.2 & 63.0 & 67.4 & 4.3 & 0.0 to 10.0 & [0.4, 9.0] & [2.6, 6.3] & 1.4 [-0.7, 3.6] \\
Decide-once, merge & 73.8 & 70.0 to 77.8 & 70.4 & 74.4 & 3.5 & 2.2 to 5.6 & [0.5, 6.7] & [2.6, 4.3] & 2.5 [-1.2, 6.5] \\
\bottomrule
\end{tabular}
\end{table*}

\section{Published Rules and Corruptions}
\label{app:sota}

Table~\ref{tab:sotafull} gives the calm accuracy and the Worst Acc of InfiniPot-V and Nexus Sampling at both budgets on the 30 test streams, Table~\ref{tab:sotaci} gives the paired intervals of every quota against InfiniPot-V at the 8 s budget, and Table~\ref{tab:corrupt} gives the accuracy of every policy under the two corruptions at the 8 s budget, which Table~\ref{tab:sota} in the main text averages over the three base policies. Every quota leads Nexus Sampling in Worst Acc and under the corruptions on all three models, and the lower end of every interval is at least 11.1\%. Under the corruptions, the quotas lead Nexus Sampling by 38.6\% to 42.8\% on Qwen2.5-VL-7B and by 27.2\% to 35.0\% on LLaVA-OneVision-7B, and on LLaVA-OneVision-7B, the intervals against InfiniPot-V lie above zero for attention and for equal spans with merge. Full memory loses at most 2.8\% under either corruption, so both segments leave the content of the earlier clips intact and act on the memory alone. On Qwen2.5-VL-3B, the quotas built on attention and merge lead InfiniPot-V by 8.6\% to 15.8\% in mean accuracy under the corruptions, with every interval above zero, while those built on value trail it by 3.9\% to 8.6\%, as on the natural stressors. On Qwen2.5-VL-7B and LLaVA-OneVision-7B, the quotas change value and merge by -3.3\% to 3.3\%, with every interval but that of equal spans with value on LLaVA-OneVision-7B including zero, and they raise attention by 25.8\% to 38.6\%, so the corruptions harm only the base policy that already fails on these two models under the natural stressors.

\section{Adaptive Adversary}
\label{app:adaptive}

Table~\ref{tab:mech} decomposes every attack of Table~\ref{tab:adaptive} in the main text into the change in the Res of the three earlier clips, the total drop in Early Acc from the calm control, the drop of full memory on the same selected segments, and the excess, which is the difference of the last two with its paired interval. Keyed rows average over the defender keys, and the secret-key rows also average over the attacker keys. The selected segments differ between defenses, so the full-memory drop differs between rows. Under the answer attack, the Res of the earlier clips hardly moves, since this attack seeks wrong answers instead of evictions. Under the token attack, the Res moves most for the base policies, by up to 7.1\% with value, and at most 2.0\% under every quota.

Table~\ref{tab:pairs} reports the Early Acc against the answer attack over the nine pairs of three defender keys and three attacker keys. The column Averaged is an adversary who averages its score over its three keys, and Gap is the secret-key accuracy minus the leaked-key accuracy. The two intervals of the gap differ in what they resample. Streams and keys resamples both the 30 streams and the key pairs, while Keys only resamples the key pairs with the streams fixed. The second interval is much narrower, so most of the uncertainty comes from the streams and little from the choice of keys. No single pair shows a clear loss, and the range starts at 0.0\% for decide-once with attention and with value.

Figure~\ref{fig:adaptive} in the main text shows how the Early Acc under the answer attack falls as the adversary searches more choices. The first three panels use the 32 choices on all 30 streams, and the last panel uses 64 choices on 15 streams with one key pair, so its values at equal search sizes differ from those of the first panel. Each point averages random subsets of the choices, and each choice costs the adversary one simulated pass of the stream. With a single choice, a leaked and a secret key give nearly the same accuracy, since there is nothing to select, and the gap of KeyMem with attention reaches 4.9\% at 8 choices. With 64 choices, a secret key keeps 60.0\% and a leaked key 46.7\%. On this key pair, decide-once scores 60.0\% with a secret key and 64.4\% with a leaked key, a difference of -4.4\% [-11.1, 0.0], and equal spans scores 73.3\%, cap only 46.7\% and full memory 60.0\%. Decide-once keeps 1.5\%, 1.1\% and 0.8\% more of the earlier tokens than KeyMem in calm streams with attention, value and merge, because a token that survives one round keeps its noise and tends to survive the next.

Table~\ref{tab:replay} reports two controls. The replay picks, on every stream, the segment that gives full memory the lowest answer margin and plays this one segment against every defense, so no defense is attacked with a segment chosen for it. Against this segment, equal spans with attention leads KeyMem by 14.1\% [6.3, 21.5] and decide-once by 8.1\% [2.2, 13.7], and the differences from equal spans with value and merge all include zero. The eviction-seeking attack keeps only segments that leave the correctness of full memory on all three questions equal to that of the calm control, which are 24.6 of the 32 on average, and it minimizes the surviving tokens among them, so that direct interference is as small as we can select for. Its drops are small under every quota, and only the two base policies show an interval above zero. This attack cannot rule out damage from a better search, since the pool holds natural footage only.

\begin{table}[t]
\centering
\footnotesize
\caption{Replay and eviction-seeking controls on Qwen2.5-VL-7B at $B=512$, with $^{*}$ for an interval above zero.}
\label{tab:replay}
\setlength{\tabcolsep}{2pt}
\begin{tabular}{@{}lcccccc@{}}
\toprule
\multirow{2}{*}{Defense} & \multicolumn{3}{c}{Replay, Early Acc} & \multicolumn{3}{c}{Eviction-seeking, drop} \\
\cmidrule(lr){2-4}\cmidrule(lr){5-7}
 & Attention & Value & Merge & Attention & Value & Merge \\
\midrule
Base policy & 28.9 & 65.6 & 81.1 & 2.2 & 14.4$^{*}$ & 6.7$^{*}$ \\
Equal spans & 81.1 & 70.0 & 80.0 & 1.1 & 3.3 & 1.1 \\
Cap only & 63.3 & 68.9 & 81.1 & 2.2 & 3.3 & 4.4 \\
KeyMem & 67.0 & 68.9 & 77.4 & -1.1 & 0.0 & 1.1 \\
Decide-once & 73.0 & 73.7 & 78.1 & -5.6 & 3.3 & 0.0 \\
\bottomrule
\end{tabular}
\end{table}

\begin{table}[t]
\centering
\footnotesize
\caption{Early accuracy on 15 streams at $B=1024$.}
\label{tab:b1024}
\setlength{\tabcolsep}{3pt}
\begin{tabular}{@{}lcccccc@{}}
\toprule
\multirow{2}{*}{Defense} & \multicolumn{3}{c}{Attention} & \multicolumn{3}{c}{Value} \\
\cmidrule(lr){2-4}\cmidrule(lr){5-7}
 & Calm & Answer & Excess & Calm & Answer & Excess \\
\midrule
Equal spans & 86.7 & 73.3 & -4.4 & 86.7 & 77.8 & -4.4 \\
Cap only & 77.8 & 55.6 & 4.4 & 82.2 & 71.1 & -4.4 \\
KeyMem, leaked & 77.8 & 55.6 & 6.7 & 80.0 & 66.7 & 2.2 \\
KeyMem, secret & 77.8 & 68.9 & -4.4 & 80.0 & 68.9 & -2.2 \\
Decide-once, leaked & 77.8 & 64.4 & 0.0 & 88.9 & 77.8 & -4.4 \\
Decide-once, secret & 77.8 & 71.1 & -8.9 & 88.9 & 80.0 & -6.7 \\
\bottomrule
\end{tabular}
\end{table}

\begin{table}[t]
\centering
\small
\caption{Sensitivity of KeyMem at $B=512$.}
\label{tab:sens}
\begin{tabular}{@{}lcccc@{}}
\toprule
\multirow{2}{*}{Setting} & \multicolumn{2}{c}{Attention + KeyMem} & \multicolumn{2}{c}{Value + KeyMem} \\
\cmidrule(lr){2-3}\cmidrule(lr){4-5}
 & Calm & Text & Calm & Text \\
\midrule
Default & 83.3 & 80.0 & 84.4 & 86.7 \\
$\tau=0.1$ & 81.1 & 84.4 & 84.4 & 80.0 \\
$\tau=1.5$ & 85.6 & 87.8 & 84.4 & 87.8 \\
$\rho=0.125$ & 86.7 & 86.7 & 84.4 & 83.3 \\
$\rho=0.25$ & 78.9 & 78.9 & 87.8 & 84.4 \\
$\rho=0.5$ & 57.8 & 65.6 & 81.1 & 65.6 \\
$W=5$ s & 76.7 & 74.4 & 80.0 & 82.2 \\
$W=20$ s & 60.0 & 60.0 & 75.6 & 67.8 \\
$R=1$ & 82.2 & 85.6 & 84.4 & 86.7 \\
$R=32$ & 84.4 & 87.8 & 80.0 & 86.7 \\
\bottomrule
\end{tabular}
\end{table}

\begin{table*}[t]
\centering
\footnotesize
\caption{Accuracy on the 200 s streams by evidence age, with residency in parentheses.}
\label{tab:longtext}
\begin{tabular}{@{}lcccccccc@{}}
\toprule
\multirow{2}{*}{Policy} & \multicolumn{4}{c}{Text segment} & \multicolumn{4}{c}{Calm control} \\
\cmidrule(lr){2-5}\cmidrule(lr){6-9}
 & All & 0--40 s & 40--100 s & ${>}$100 s & All & 0--40 s & 40--100 s & ${>}$100 s \\
\midrule
Full memory & 74.1 & 81.7 (100.0) & 70.0 (100.0) & 73.3 (100.0) & 73.0 & 81.7 (100.0) & 67.8 (100.0) & 72.5 (100.0) \\
\midrule
Attention & 25.9 & 46.7 (19.5) & 18.9 (0.2) & 20.8 (0.1) & 25.6 & 45.0 (19.5) & 17.8 (0.3) & 21.7 (0.1) \\
\quad + Equal spans & 46.3 & 40.0 (4.9) & 27.8 (0.5) & \textbf{63.3} (5.0) & 45.2 & 36.7 (4.9) & 28.9 (0.5) & \textbf{61.7} (5.1) \\
\quad + Cap only & 48.1 & \textbf{80.0} (12.2) & 45.6 (3.1) & 34.2 (1.2) & 49.3 & \textbf{76.7} (12.3) & 48.9 (3.2) & 35.8 (1.1) \\
\quad + KeyMem & 48.1 & 78.3 (12.7) & 44.4 (3.5) & 35.8 (0.8) & 49.6 & \textbf{76.7} (12.8) & 47.8 (3.5) & 37.5 (0.8) \\
\quad + Decide-once & \textbf{58.1} & 73.3 (7.1) & \textbf{52.2} (3.7) & 55.0 (2.9) & \textbf{58.1} & 73.3 (7.1) & \textbf{53.3} (3.7) & 54.2 (3.0) \\
\midrule
Value & 43.0 & 36.7 (1.3) & 35.6 (1.0) & 51.7 (1.4) & 55.2 & 51.7 (3.2) & 51.1 (3.4) & \textbf{60.0} (4.5) \\
\quad + Equal spans & 55.6 & 60.0 (3.9) & 50.0 (3.9) & 57.5 (3.8) & 49.3 & 46.7 (3.9) & 44.4 (3.9) & 54.2 (4.1) \\
\quad + Cap only & 55.6 & 56.7 (2.8) & 53.3 (2.7) & 56.7 (3.0) & 55.2 & 58.3 (3.6) & 48.9 (3.7) & 58.3 (4.2) \\
\quad + KeyMem & 46.3 & \textbf{78.3} (9.5) & 40.0 (2.6) & 35.0 (1.0) & 52.6 & \textbf{75.0} (9.8) & 48.9 (3.3) & 44.2 (2.1) \\
\quad + Decide-once & \textbf{61.9} & 73.3 (4.6) & \textbf{58.9} (4.6) & \textbf{58.3} (3.2) & \textbf{58.9} & 70.0 (4.8) & \textbf{56.7} (4.9) & 55.0 (3.3) \\
\midrule
Merge & 52.6 & 63.3 (3.6) & 42.2 (2.1) & 55.0 (3.9) & 55.9 & 66.7 (4.0) & 45.6 (2.8) & 58.3 (4.9) \\
\quad + Equal spans & 64.8 & 60.0 (4.0) & 60.0 (3.2) & \textbf{70.8} (4.3) & \textbf{59.6} & 58.3 (3.9) & \textbf{55.6} (3.2) & \textbf{63.3} (4.5) \\
\quad + Cap only & 57.8 & 66.7 (4.0) & 50.0 (2.6) & 59.2 (4.0) & 58.1 & 61.7 (4.6) & 48.9 (2.9) & \textbf{63.3} (4.5) \\
\quad + KeyMem & 57.8 & \textbf{80.0} (10.4) & 48.9 (3.2) & 53.3 (1.6) & 57.0 & \textbf{80.0} (10.5) & 51.1 (3.3) & 50.0 (1.8) \\
\quad + Decide-once & \textbf{65.2} & 70.0 (4.6) & \textbf{61.1} (4.6) & 65.8 (3.5) & \textbf{59.6} & 65.0 (4.6) & \textbf{55.6} (4.6) & 60.0 (3.6) \\
\bottomrule
\end{tabular}
\end{table*}

Table~\ref{tab:b1024} repeats the adversary at $B=1024$ on the first 15 test streams with one key pair. Equal spans again gives the best answer-attack accuracy with attention, and decide-once with a secret key gives the best with value. No excess has an interval above zero at this budget, and with 15 streams, the intervals are wide, so this setting supports the direction of the main comparison and does not fix its size.

\section{Sensitivity}
\label{app:sens}

Table~\ref{tab:sens} varies one hyperparameter of KeyMem at a time from the default of $\tau=0.5$, $\rho=0.1875$, $W=10$ s and $R=8$. It reports the Acc over all clips on the first 15 test streams of Qwen2.5-VL-7B at $B=512$, with the calm control and the text segment, which is the stressor that breaks the value policy. The cap ratio and the bucket length matter most, because together they set how many tokens per second of video any part of the stream may keep. With $\rho=0.5$, a single 10 s bucket may hold half of the memory, and the accuracy of attention falls to 57.8\% in calm streams. With $W=20$ s, the 80 s stream spans only four buckets whose caps add up to 384 tokens, so a quarter of the budget stays empty, and with $W=5$ s, a 20 s segment may again fill most of the memory. A smaller cap helps on these streams, and $\rho=0.125$ reaches 86.7\% with attention in both conditions, slightly above the $\rho=0.1875$ that the validation streams select. The temperature and the epoch length matter less, and since the chunk index already renews the draw in every round, $R$ changes only the seed. No setting of $\tau$ from 0.1 to 1.5 or of $R$ from 1 to 32 falls more than 6.7\% below the default, and several settings exceed it, so the draw can be tuned for accuracy without giving up the protection of the cap. Appendix~\ref{app:validation} lists the validation results that select the values of $\tau$ and $\rho$.

\section{Long Streams}
\label{app:long}

The age of a clip is the time between the end of the clip and the end of the stream, when all 18 questions are asked. Each stress and policy holds 60 questions younger than 40 s, 90 between 40 s and 100 s and 120 older than 100 s, and bold marks the best rule per base policy and column. Table~\ref{tab:longtext} adds the Res, the middle ages and the calm streams to Table~\ref{tab:long} in the main text, and the same pattern holds without a stressor. KeyMem leads on the newest clips with every base policy, equal spans keeps the most old evidence with attention, and decide-once matches or leads the other quotas over all clips. Full memory scores lower on the middle ages than on the oldest clips, so the difficulty of the questions also varies with the position of a clip, and the comparison between policies at the same age is the meaningful one. On the text streams, decide-once keeps 2.9\% to 3.5\% of the tokens of clips older than 100 s. Over all clips, it leads equal spans by 11.9\% [5.2, 18.1], 6.3\% [0.0, 11.9] and 0.4\% [-5.2, 6.3], and it gives up 5.0\%, 5.0\% and 10.0\% to KeyMem on the newest clips, where only the last interval, [0.0, 21.7], does not extend below zero, so only merge pays a clear price. Table~\ref{tab:long60} reports the long-duration content on the same 15 streams. The 60 s text segment follows the third clip, and it lengthens the streams to 240 s. The overlay uses the calm streams and adds the caption bar, the ticker and the clock to every frame, including the frames of the clips, which is why full memory also falls to 64.4\%. Value then scores between 18.1\% and 24.1\% with every rule, close to chance, since a temporal rule cannot repair a failure of content that is spread over all of time. On clips older than 100 s, decide-once leads KeyMem by 15.0\% [5.8, 24.2], 35.0\% [20.8, 48.3] and 17.5\% [4.2, 30.8] with attention, value and merge on the 60 s segment, and by 13.3\% [5.0, 23.3], 6.7\% [-0.8, 15.8] and 20.8\% [11.7, 29.2] with the overlay.

\begin{table}[t]
\centering
\footnotesize
\caption{Accuracy by evidence age (s) with a 60 s text segment and with a text overlay on the whole stream.}
\label{tab:long60}
\setlength{\tabcolsep}{3pt}
\begin{tabular}{@{}lcccccc@{}}
\toprule
\multirow{2}{*}{Policy} & \multicolumn{3}{c}{60 s text segment} & \multicolumn{3}{c}{Whole-stream overlay} \\
\cmidrule(lr){2-4}\cmidrule(lr){5-7}
 & All & 0--40 & ${>}$100 & All & 0--40 & ${>}$100 \\
\midrule
Full memory & 73.3 & 85.0 & 70.0 & 64.4 & 75.0 & 58.3 \\
\midrule
Attention & 24.8 & 41.7 & 21.7 & 20.0 & 40.0 & 14.2 \\
\quad + Equal spans & 34.8 & 40.0 & \textbf{43.3} & 27.0 & 38.3 & 32.5 \\
\quad + Cap only & 47.4 & \textbf{86.7} & 30.8 & 34.8 & 71.7 & 21.7 \\
\quad + KeyMem & 45.2 & 83.3 & 28.3 & 37.4 & \textbf{73.3} & 26.7 \\
\quad + Decide-once & \textbf{51.5} & 75.0 & \textbf{43.3} & \textbf{40.7} & 65.0 & \textbf{40.0} \\
\midrule
Value & 18.9 & 28.3 & 20.8 & 18.1 & 20.0 & 14.2 \\
\quad + Equal spans & 52.2 & 53.3 & 53.3 & 18.5 & 26.7 & 16.7 \\
\quad + Cap only & 24.1 & 38.3 & 20.0 & 18.9 & 25.0 & 13.3 \\
\quad + KeyMem & 40.4 & \textbf{75.0} & 23.3 & 19.3 & \textbf{28.3} & 17.5 \\
\quad + Decide-once & \textbf{64.4} & 71.7 & \textbf{58.3} & \textbf{24.1} & 26.7 & \textbf{24.2} \\
\midrule
Merge & 53.3 & 55.0 & 56.7 & 38.1 & 51.7 & 38.3 \\
\quad + Equal spans & 56.3 & 65.0 & \textbf{58.3} & 38.9 & 46.7 & 35.0 \\
\quad + Cap only & 52.6 & 60.0 & 56.7 & 35.2 & 55.0 & 32.5 \\
\quad + KeyMem & 51.5 & \textbf{78.3} & 38.3 & 34.4 & \textbf{63.3} & 22.5 \\
\quad + Decide-once & \textbf{59.6} & 66.7 & 55.8 & \textbf{48.1} & 60.0 & \textbf{43.3} \\
\bottomrule
\end{tabular}
\end{table}

\begin{table*}[t]
\centering
\footnotesize
\caption{Accuracy on the text streams of Qwen2.5-VL-3B at 200 s and Qwen2.5-VL-7B at 400 s by evidence age (s).}
\label{tab:longp1}
\begin{tabular}{@{}lccccccccc@{}}
\toprule
\multirow{2}{*}{Policy} & \multicolumn{4}{c}{Qwen2.5-VL-3B, 200 s} & \multicolumn{5}{c}{Qwen2.5-VL-7B, 400 s} \\
\cmidrule(lr){2-5}\cmidrule(lr){6-10}
 & All & 0--40 & 40--100 & ${>}$100 & All & 0--40 & 40--100 & 100--200 & ${>}$200 \\
\midrule
Full memory & 78.1 & 81.7 & 76.7 & 77.5 & 60.7 & 71.7 & 60.0 & 52.7 & 63.0 \\
\midrule
Attention & 44.8 & 75.0 & 41.1 & 32.5 & 21.9 & 46.7 & 20.0 & 20.7 & 17.8 \\
\quad + Equal spans & 45.9 & 46.7 & 27.8 & 59.2 & 32.3 & 35.0 & 23.3 & 18.0 & \textbf{42.6} \\
\quad + Cap only & \textbf{69.6} & \textbf{86.7} & \textbf{66.7} & \textbf{63.3} & 26.0 & 71.7 & 32.2 & 17.3 & 18.5 \\
\quad + KeyMem & 61.1 & 85.0 & 61.1 & 49.2 & 27.4 & \textbf{73.3} & 38.9 & 18.7 & 18.1 \\
\quad + Decide-once & 65.6 & 80.0 & 60.0 & 62.5 & \textbf{37.0} & 58.3 & \textbf{46.7} & \textbf{36.0} & 29.6 \\
\midrule
Value & 38.1 & 38.3 & 38.9 & 37.5 & 28.8 & 26.7 & 28.9 & 30.0 & 28.5 \\
\quad + Equal spans & 41.5 & 36.7 & 46.7 & \textbf{40.0} & 38.2 & 55.0 & 42.2 & 34.0 & 35.6 \\
\quad + Cap only & 38.1 & 36.7 & 41.1 & 36.7 & 37.0 & 35.0 & 34.4 & 38.0 & 37.8 \\
\quad + KeyMem & 38.9 & \textbf{60.0} & 43.3 & 25.0 & 30.9 & \textbf{73.3} & 50.0 & 26.7 & 17.4 \\
\quad + Decide-once & \textbf{47.8} & 53.3 & \textbf{57.8} & 37.5 & \textbf{49.5} & 56.7 & \textbf{53.3} & \textbf{47.3} & \textbf{47.8} \\
\midrule
Merge & 56.7 & 65.0 & 53.3 & 55.0 & 42.8 & 40.0 & 34.4 & 44.0 & 45.6 \\
\quad + Equal spans & \textbf{57.4} & 58.3 & 56.7 & \textbf{57.5} & \textbf{45.3} & 48.3 & 38.9 & \textbf{46.0} & 46.3 \\
\quad + Cap only & 56.7 & 65.0 & 55.6 & 53.3 & 43.2 & 45.0 & 37.8 & 39.3 & \textbf{46.7} \\
\quad + KeyMem & 53.7 & \textbf{86.7} & 52.2 & 38.3 & 36.8 & \textbf{73.3} & \textbf{57.8} & 32.7 & 24.1 \\
\quad + Decide-once & 54.8 & 58.3 & \textbf{63.3} & 46.7 & 43.7 & 51.7 & 41.1 & \textbf{46.0} & 41.5 \\
\bottomrule
\end{tabular}
\end{table*}

Table~\ref{tab:longp1} extends the long streams to a second model and a longer horizon, with the text segment after the third clip, 15 test streams each and $B=512$. The 3B streams have the 18 clips of the 200 s streams, and the 400 s streams have 38 clips, so the budget holds 2\% of them. LLaVA-OneVision encodes 392 tokens per second with one position per token, so a 200 s stream would need about 78k positions, beyond the 32k context of its language model, and we do not run it here. On the 3B model, the pattern of Table~\ref{tab:longtext} holds. On clips older than 100 s, KeyMem trails equal spans by 10.0\% [1.7, 18.3], 15.0\% [6.7, 22.5] and 19.2\% [13.3, 25.8] with attention, value and merge, while decide-once differs from equal spans there by 3.3\% [-5.0, 11.7], -2.5\% [-11.7, 7.5] and -10.8\% [-16.7, -3.3]. Equal spans with attention again neglects the newest clips and scores 46.7\% on them against 86.7\% for cap only, and the calm 3B streams give a similar order. At 400 s, full memory falls to 60.7\% and the base policies to between 21.9\% and 42.8\%. On clips older than 200 s, KeyMem trails cap only by 0.4\% [-4.1, 5.2], 20.4\% [14.8, 26.7] and 22.6\% [14.8, 28.9], and over all clips, decide-once leads equal spans by 4.7\% [-0.5, 9.8], 11.2\% [6.0, 16.3] and -1.6\% [-5.6, 2.3]. Attention fails at this horizon with every rule, as none exceeds 37.0\% overall.

We also run the adaptive adversary of Section~\ref{sec:adversary} in the main text on 15 of the 200 s text streams of Section~\ref{sec:long}, with one key pair and no full-memory control. With attention, equal spans and decide-once with a secret key both keep 64.4\%, a difference of 0.0\% [-15.6, 15.6]. The segments use the same stream positions as the text segment. These streams use almost every class in their questions, so the candidates exclude only the classes of the stream and of the three earlier questions. With the answer objective and merge, equal spans scores 62.2\%, cap only 80.0\%, KeyMem 46.7\% with a leaked key and 77.8\% with a secret key, and decide-once 71.1\% and 75.6\%, so decide-once with a secret key leads equal spans by 13.3\% [-2.2, 28.9]. With the token objective, the attack is again weaker for decide-once, which scores 66.7\% with a secret key and 60.0\% with a leaked key with attention, and 77.8\% and 75.6\% with merge. KeyMem scores 33.3\% with attention under this objective, 42.2\% [26.7, 60.0] below equal spans, and decide-once trails equal spans by 8.9\% [-4.4, 22.2]. These results rest on one key pair and 15 streams, so they show the direction of the trade-off between the rules but leave its size open for a larger study.

\section{Summary of the Temporal Rules}
\label{app:decision}

Table~\ref{tab:decision} summarizes the four temporal rules of Section~\ref{sec:decision} on Qwen2.5-VL-7B at $B=512$, with each number averaged over the three base policies. The columns deliberately separate empirical accuracy, long-range retention and formal guarantees. Worst Acc comes from the 80 s streams, Acc ${>}$100 s from the 200 s text streams and the answer attack from Table~\ref{tab:adaptive}, using a secret key where applicable. Fixed Bound records whether every segment of known duration has a content-independent occupancy limit at every elapsed time, while Exact Attack records whether an adversary without the key can simulate the retention decision exactly. Equal spans best preserves old evidence, the deterministic cap supplies the occupancy guarantee without key management, and keyed rules add unpredictability rather than a uniform accuracy gain. On StreamingBench, decide-once shows no clear gain over its base policies, so its advantage for old evidence is supported specifically by the 200 s and 400 s stress streams.

\begin{table}[t]
\centering
\footnotesize
\caption{Summary of the four temporal rules.}
\label{tab:decision}
\setlength{\tabcolsep}{2.2pt}
\begin{tabular}{@{}lccccc@{}}
\toprule
Rule & \stk{Worst\\Acc} & \stk{Acc\\${>}$100 s} & \stk{Answer\\attack} & \stk{Fixed\\bound} & \stk{Exact\\attack} \\
\midrule
Equal spans & 80.4 & \textbf{63.9} & \textbf{68.5} & No & Yes \\
Cap only & 79.1 & 50.0 & 60.0 & Yes & Yes \\
KeyMem & 79.6 & 41.4 & 63.3 & Yes & No \\
Decide-once & \textbf{80.9} & 59.7 & 67.4 & Yes & No \\
\bottomrule
\end{tabular}
\end{table}

\section{StreamingBench Breakdown}
\label{app:sbench}

\begin{table}[t]
\centering
\footnotesize
\caption{Accuracy and circular accuracy on StreamingBench by temporal clue and by query time.}
\label{tab:sbench}
\setlength{\tabcolsep}{1.3pt}
\begin{tabular}{@{}lccccccc@{}}
\toprule
Policy & All & Circular & \stk{Earlier\\moment} & \stk{Current\\moment} & 0--60 s & 60--180 s & ${>}$180 s \\
\midrule
Full memory & 69.3 & 69.8 & 70.0 & 69.2 & 81.4 & 68.9 & 65.5 \\
Sliding window & 64.5 & 65.7 & 69.3 & 63.3 & 80.0 & 65.3 & 58.9 \\
\midrule
Attention & 66.0 & 67.1 & 73.6 & 64.2 & 80.0 & 69.5 & 59.9 \\
\enspace + Equal spans & \textbf{67.8} & 67.3 & 71.4 & 67.0 & 80.7 & 67.7 & \textbf{63.5} \\
\enspace + Cap only & 66.0 & 67.4 & 73.6 & 64.2 & 79.3 & 70.1 & 59.9 \\
\enspace + KeyMem & 65.6 & 66.4 & 74.3 & 63.5 & 79.3 & 67.7 & 60.1 \\
\enspace + Decide-once & 65.5 & 66.7 & 69.3 & 64.5 & 79.3 & 65.3 & 60.9 \\
\midrule
Value & 63.0 & 62.7 & 65.7 & 62.3 & 80.7 & 62.3 & 57.2 \\
\enspace + Equal spans & 63.1 & 62.7 & 65.7 & 62.5 & 80.7 & 64.7 & 56.5 \\
\enspace + Cap only & 61.7 & 61.7 & 63.6 & 61.3 & 80.0 & 62.3 & 55.3 \\
\enspace + KeyMem & \textbf{66.2} & 68.0 & 70.7 & 65.1 & 80.0 & 68.3 & \textbf{60.6} \\
\enspace + Decide-once & 64.6 & 64.8 & 67.1 & 64.0 & 80.0 & 64.1 & 59.7 \\
\midrule
Merge & 65.5 & 65.7 & 66.4 & 65.2 & 80.7 & 60.5 & 62.3 \\
\enspace + Equal spans & 66.0 & 65.7 & 67.9 & 65.6 & 80.7 & 62.3 & \textbf{62.6} \\
\enspace + Cap only & 65.5 & 65.6 & 67.1 & 65.1 & 81.4 & 62.3 & 61.4 \\
\enspace + KeyMem & \textbf{67.3} & 67.4 & 72.1 & 66.1 & 81.4 & 68.3 & 62.1 \\
\enspace + Decide-once & 65.5 & 65.9 & 67.9 & 64.9 & 81.4 & 64.1 & 60.6 \\
\bottomrule
\end{tabular}
\end{table}

The 140 questions about an earlier moment come closest to delayed evidence. On them, full memory reaches 70.0\% and the sliding window 69.3\%, so this subset rewards memory only weakly. KeyMem changes the base policies by 0.7\% [-2.3, 3.9], 5.0\% [-0.7, 10.7] and 5.7\% [0.7, 10.7] for attention, value and merge, and equal spans by -2.1\% [-7.5, 3.1], 0.0\% [-4.5, 4.4] and 1.4\% [-2.6, 5.6], so most of these comparisons remain unresolved. Decide-once changes the base policies by -4.3\% [-9.0, 0.0], 1.4\% [-4.0, 6.5] and 1.4\% [-3.4, 6.2] on these questions, and its one interval against KeyMem that excludes zero is -5.0\% [-9.2, -1.3] with attention. After 180 s, it changes the base policies by 1.0\% [-2.1, 4.1], 2.4\% [-1.2, 6.1] and -1.7\% [-4.8, 1.5]. Over all questions, every interval of decide-once against KeyMem includes zero, at -0.1\% [-2.1, 1.8], -1.5\% [-3.9, 1.0] and -1.8\% [-4.1, 0.6]. Table~\ref{tab:sbench} also reports the circular accuracy, the accuracy by the temporal clue of each question and the accuracy by query time. The 140 questions about an earlier moment and the 581 about the current moment follow the labels of the benchmark, and the three time ranges hold 140, 167 and 414 questions. Circular evaluation averages the score of each option over the four cyclic orders, as on our coverage benchmark, and changes no conclusion. Within the first 60 s, the stream still fits the memory almost entirely, and every policy scores between 79.3\% and 81.4\%. The gap to full memory opens after 60 s and is widest after 180 s, which is where the quotas differ most. Questions about an earlier moment are answered about as well as questions about the current moment, so the label alone does not mark evidence that has left a memory of 28 s, and the query time separates the policies more clearly.

\section{Implementation Details}
\label{app:impl}

\noindent\textbf{Streams and Questions.} Frames are resized to $224\times224$, and the variants of a stream differ only in the stress segment, so every accuracy over all clips averages 180 questions and every accuracy over the three clips before the segment averages 90. One option of each question is the class of its clip. A blind control that sees no video scores between 16.1\% and 19.4\% in calm streams, near the chance level of 20\%.

\noindent\textbf{Stress Segments.} We rank the test clips of at least 10 s by the mean absolute difference between consecutive grayscale frames and call the lowest tenth calm and the highest tenth busy. Text, flicker and shake reuse the footage of the calm control. The text overlay changes every few frames, flicker multiplies the brightness of alternate frames by 1.45 and 0.55, and shake rotates each frame by up to 8 degrees, shifts it by up to 18 pixels and zooms it by 1.3. The Gaussian noise has a standard deviation of 12\% of the pixel range, and the Gaussian blur has a radius of 3 pixels. Adversarial pixel perturbations fall outside the threat model, since the adversary cannot edit pixels after capture.

\noindent\textbf{Models and Policies.} Qwen2.5-VL turns every chunk into 256 visual tokens with multimodal rotary positions, and LLaVA-OneVision turns it into 196 tokens per frame plus a newline token. Each token keeps its position after eviction, and the two budgets hold 10\% and 20\% of the stream. The attention policy simplifies H2O~\cite{zhang2023h2o} and SnapKV~\cite{li2024snapkv}. The value policy multiplies the norm of the last-layer value vector by one minus half its cosine similarity with the previous token, which simplifies InfiniPot-V~\cite{kim2025infinipot}, and the merge policy keeps the tokens that differ most from their predecessor. KeyMem uses one secret key on the coverage streams, and Table~\ref{tab:keys} adds two keys. We select $\tau$ and then $\rho$ on the validation streams, which share no clip with the test streams. Cap only keeps the rank normalization and the hard cap of KeyMem. StreamMem scores tokens with the attention of generic queries, which our attention policy approximates, and ReST-KV reconstructs the output of each layer, which one retention set for all layers cannot express, so we leave both out of our comparison with published rules.

\noindent\textbf{Adversaries.} Of the 16 candidates of a stream, eleven come from one of the five stress types and five are spliced from random types, and each candidate starts at the grid-aligned start after the third clip or 5 s later. The answer objective takes the margin of the correct option over the best wrong option, which uses the model as an oracle, and every choice is replayed under every defense. Two further settings use the first 15 test streams and one key pair, at $B=1024$ and with 64 choices, and Appendix~\ref{app:first} adds a weaker adversary that counts tokens over 16 natural choices.

\noindent\textbf{Software and GPUs.} All runs use Hugging Face Transformers 5.19 in bfloat16 on one GPU per process, with eager attention for identical numerics across policies. With an unlimited budget, our harness reproduces one offline forward pass over the whole prompt (Appendix~\ref{app:validation}). LLaVA-OneVision prefills each chunk in causal pieces of two frames to bound the memory of eager attention, which leaves the attention map of a single pass unchanged.

\noindent\textbf{Keyed Generator.} The uniform variables of the keyed draw come from a PyTorch generator whose seed is the first 8 bytes of the SHA-256 hash of the key, the epoch $\lfloor t/R\rfloor$ and the chunk index $t$. The key itself stays fixed for the whole stream. The decide-once variant instead hashes the key with the identifier of each token, so a token receives one Gumbel perturbation when it arrives and keeps it in every later round, while the ranks and the cap work as in KeyMem.

\noindent\textbf{Quota Baselines.} Equal spans divides the time between the oldest and the newest candidate into 8 spans of equal length. It visits the spans from the one with the fewest candidates to the one with the most, gives each the remaining budget divided by the number of spans that are left, and keeps the highest-scoring tokens of a span that holds more candidates than its share. Cap only runs the code of KeyMem with $\tau=0.01$, which makes the largest weight ratio $e^{100}$ and the draw a top-$k$ selection in practice. The shifted grid adds 5 s to the timestamps of the candidates before the cap assigns them to its buckets of $W$ seconds.

\noindent\textbf{Published Rules.} InfiniPot-V follows Algorithm 1 of its paper, keeps the newest frames up to 12.5\% of the budget and gives its TaR score a share of $\alpha=0.5$ of the budget. Its TaR score is the negative cosine similarity between the key of a past token and the key at the same patch position in each recent frame, averaged over the heads, the recent frames and all layers. Its VaN score is the value norm, averaged over the heads and over all layers after dividing each layer by its mean, and smoothed with a $3\times3$ average pool over the patch grid of its frame, since the paper does not publish the thresholds of its adaptive kernel. On LLaVA-OneVision-7B, this grid holds the 14$\times$14 tokens of a frame, and the newline token of each chunk is never removed, as under every other rule. The paper compresses each layer separately once the cache reaches a larger budget, while our harness keeps one retention set for all layers and compresses to the budget after every chunk, as it does for every other rule. Nexus Sampling follows Equations 1 to 6 of its paper with blocks of 32 tokens, a bridge walk of 3 steps with weight $\lambda=0.5$, a tie-break ramp of $10^{-6}$ and 5 uniform draws per priority. Its observation window is the set of queries of the newest chunk, which our attention policy also uses, since a stream carries no text query, and it protects no recency window, because the system and text tokens are never removed.

\noindent\textbf{Adaptive Attack.} For every stream, the harness runs the memory once up to the end of the third clip, stores a copy of the cache and of the encoder state, and restores this copy before each of the 33 continuations, which are the calm control and the 32 choices. Each continuation records the margin, the correctness and the residency of the three earlier questions, so an attack under any objective and any search size reads its outcome from the same matrix. The secret-key attacker runs the same continuations under its own key, picks the choice that is worst under that key, and is then scored under the defender's key. Each smaller search size averages 50 random subsets of the choices per stream.

\noindent\textbf{StreamingBench Protocol.} We use the released parts 1 to 50, 51 to 100 and 201 to 250 of the benchmark. Each video is decoded at 1 frame per second and $336\times336$ pixels, which gives 72 visual tokens per second, and every question time is rounded down to an even number of frames, because Qwen2.5-VL merges two consecutive frames into one temporal patch. The harness encodes the video once with chunk boundaries every 8 s and at every question time, and it replays the same token chunks for every policy, so all policies see identical visual tokens. A question is answered from a copy of the memory at its time, with the four options in the order of the benchmark and the instruction to answer with the letter of an option, and the predicted option is the one whose letter receives the highest log-probability.

\noindent\textbf{Bootstrap and Compute.} Every interval is the central 95\% range of paired bootstrap resamples. We draw 10,000 resamples of streams for the coverage benchmark, the adaptive adversary and the long streams, and 10,000 resamples of videos weighted by their number of questions for StreamingBench. All runs use NVIDIA RTX PRO 5000 GPUs with 48 GB of memory. One defense on one stream of the adaptive adversary, with all 33 continuations, takes about 75 s on Qwen2.5-VL-7B, and one StreamingBench video takes 1.5 to 3 minutes.

\section{Overhead}
\label{app:overhead}

The retention rule of KeyMem needs one sort for the ranks, one pass over the buckets and one top-$k$ over the perturbed scores, all over at most $B+c_t$ scalars per chunk, and it reads no tensor that the base policy does not already compute. On one CPU core, a retention decision takes 0.3 ms for Qwen2.5-VL at $B=512$ and 13.3 ms for LLaVA-OneVision at $B=3136$, against 0.03 and 0.10 ms for the top-$k$ step of the base policy. One full memory update, from the frames of a chunk to the pruned cache, takes about 110 ms on Qwen2.5-VL-7B and about 470 ms on LLaVA-OneVision with or without KeyMem, so the difference stays within the run-to-run variation of the update and adds no visible latency to the stream.

\end{document}